\pdfoutput=1

\documentclass[12pt,a4paper]{article}

\usepackage{ifthen} 
\newboolean{pdflatex}
\setboolean{pdflatex}{true} 

\newboolean{articletitles}
\setboolean{articletitles}{true} 

\newboolean{uprightparticles}
\setboolean{uprightparticles}{false} 

\newboolean{inbibliography}
\setboolean{inbibliography}{false} 

\usepackage[top=1in, bottom=1.25in, left=1in, right=1in]{geometry}

\usepackage{microtype}
\usepackage{lineno}  
\usepackage{xspace} 
\usepackage{caption} 

\usepackage{graphicx}  
\usepackage{color}
\usepackage{colortbl}
\graphicspath{{./figuresPaper/}} 
\graphicspath{{./LHCb-PAPER-2017-015-figures}} 

\usepackage{amsmath} 
\usepackage{amssymb}
\usepackage{amsfonts}
\usepackage{upgreek} 

\newcommand*\patchAmsMathEnvironmentForLineno[1]{%
\expandafter\let\csname old#1\expandafter\endcsname\csname #1\endcsname
\expandafter\let\csname oldend#1\expandafter\endcsname\csname
end#1\endcsname
 \renewenvironment{#1}%
   {\linenomath\csname old#1\endcsname}%
   {\csname oldend#1\endcsname\endlinenomath}%
}
\newcommand*\patchBothAmsMathEnvironmentsForLineno[1]{%
  \patchAmsMathEnvironmentForLineno{#1}%
  \patchAmsMathEnvironmentForLineno{#1*}%
}
\AtBeginDocument{%
\patchBothAmsMathEnvironmentsForLineno{equation}%
\patchBothAmsMathEnvironmentsForLineno{align}%
\patchBothAmsMathEnvironmentsForLineno{flalign}%
\patchBothAmsMathEnvironmentsForLineno{alignat}%
\patchBothAmsMathEnvironmentsForLineno{gather}%
\patchBothAmsMathEnvironmentsForLineno{multline}%
\patchBothAmsMathEnvironmentsForLineno{eqnarray}%
}

\usepackage{hyperref}    
\usepackage[all]{hypcap} 

\usepackage{xspace} 
\usepackage{upgreek}

\def\lhcb {\mbox{LHCb}\xspace}

\def\MagUp {\mbox{\em Mag\kern -0.05em Up}\xspace}

\ifthenelse{\boolean{uprightparticles}}%
{

 \def\Ppsi        {\ensuremath{\uppsi}\xspace}

 \def\PDelta      {\ensuremath{\Delta}\xspace}                 
 \def\PXi      {\ensuremath{\Xi}\xspace}                 
 \def\PLambda      {\ensuremath{\Lambda}\xspace}                 
 \def\PSigma      {\ensuremath{\Sigma}\xspace}                 
 \def\POmega      {\ensuremath{\Omega}\xspace}                 
 \def\PUpsilon      {\ensuremath{\Upsilon}\xspace}                 
 
 \def\PB      {\ensuremath{\mathrm{B}}\xspace}                 
                  
 \def\PD      {\ensuremath{\mathrm{D}}\xspace}

 \def\PJ      {\ensuremath{\mathrm{J}}\xspace}                 
 \def\PK      {\ensuremath{\mathrm{K}}\xspace}

 \def\Pb      {\ensuremath{\mathrm{b}}\xspace}                 
 \def\Pc      {\ensuremath{\mathrm{c}}\xspace}

 \def\Pi      {\ensuremath{\mathrm{i}}\xspace}

}
{

 \def\Ppsi        {\ensuremath{\psi}\xspace}                 
                  
 \mathchardef\PDelta="7101
 \mathchardef\PXi="7104
 \mathchardef\PLambda="7103
 \mathchardef\PSigma="7106
 \mathchardef\POmega="710A
 \mathchardef\PUpsilon="7107
                  
 \def\PB      {\ensuremath{B}\xspace}                 
                  
 \def\PD      {\ensuremath{D}\xspace}

 \def\PJ      {\ensuremath{J}\xspace}                 
 \def\PK      {\ensuremath{K}\xspace}

 \def\Pb      {\ensuremath{b}\xspace}                 
 \def\Pc      {\ensuremath{c}\xspace}

 \def\Pi      {\ensuremath{i}\xspace}

}

\makeatletter
\ifcase \@ptsize \relax
  \newcommand{\miniscule}{\@setfontsize\miniscule{4}{5}}
\or
  \newcommand{\miniscule}{\@setfontsize\miniscule{5}{6}}
\or
  \newcommand{\miniscule}{\@setfontsize\miniscule{5}{6}}
\fi
\makeatother

\DeclareRobustCommand{\optbar}[1]{\shortstack{{\miniscule (\rule[.5ex]{1.25em}{.18mm})}
  \\ [-.7ex] $#1$}}

\def\cquark    {{\ensuremath{\Pc}}\xspace}

\def\bquark    {{\ensuremath{\Pb}}\xspace}

  \def\Kbar    {{\kern 0.2em\overline{\kern -0.2em \PK}{}}\xspace}

\def\KorKbar    {\kern 0.18em\optbar{\kern -0.18em K}{}\xspace}

  \def\Dbar    {{\kern 0.2em\overline{\kern -0.2em \PD}{}}\xspace}
\def\D       {{\ensuremath{\PD}}\xspace}

\def\DorDbar    {\kern 0.18em\optbar{\kern -0.18em D}{}\xspace}
\def\Dz      {{\ensuremath{\D^0}}\xspace}

\def\Bbar    {{\ensuremath{\kern 0.18em\overline{\kern -0.18em \PB}{}}}\xspace}

\def\BorBbar    {\kern 0.18em\optbar{\kern -0.18em B}{}\xspace}

\def\jpsi     {{\ensuremath{{\PJ\mskip -3mu/\mskip -2mu\Ppsi\mskip 2mu}}}\xspace}
\def\psitwos  {{\ensuremath{\Ppsi{(2S)}}}\xspace}

  \def\Y#1S{\ensuremath{\PUpsilon{(#1S)}}\xspace}

\def\Lbar        {{\ensuremath{\kern 0.1em\overline{\kern -0.1em\PLambda}}}\xspace}
\def\LorLbar    {\kern 0.18em\optbar{\kern -0.18em \PLambda}{}\xspace}

\def\BF         {{\ensuremath{\mathcal{B}}}\xspace}

\def\BR         {\BF}

\def\to                 {\ensuremath{\rightarrow}\xspace}

\newcommand{\etot}{{\ensuremath{\varepsilon_{\mathrm{ tot}}}}\xspace}

\def\AT#1     {\ensuremath{A_{\mathrm{T}}^{#1}}\xspace}           

\def\C#1      {\ensuremath{\mathcal{C}_{#1}}\xspace}                       
\def\Cp#1     {\ensuremath{\mathcal{C}_{#1}^{'}}\xspace}                    
\def\Ceff#1   {\ensuremath{\mathcal{C}_{#1}^{\mathrm{(eff)}}}\xspace}        
\def\Cpeff#1  {\ensuremath{\mathcal{C}_{#1}^{'\mathrm{(eff)}}}\xspace}       
\def\Ope#1    {\ensuremath{\mathcal{O}_{#1}}\xspace}                       
\def\Opep#1   {\ensuremath{\mathcal{O}_{#1}^{'}}\xspace}                    

\newcommand{\tev}{\ifthenelse{\boolean{inbibliography}}{\ensuremath{~T\kern -0.05em eV}\xspace}{\ensuremath{\mathrm{\,Te\kern -0.1em V}}}\xspace}
\newcommand{\tevv}{\ensuremath{\mathrm{\,Te\kern -0.1em V}}\xspace}
\newcommand{\gev}{\ensuremath{\mathrm{\,Ge\kern -0.1em V}}\xspace}
\newcommand{\mev}{\ensuremath{\mathrm{\,Me\kern -0.1em V}}\xspace}
\newcommand{\kev}{\ensuremath{\mathrm{\,ke\kern -0.1em V}}\xspace}
\newcommand{\ev}{\ensuremath{\mathrm{\,e\kern -0.1em V}}\xspace}
\newcommand{\gevc}{\ensuremath{{\mathrm{\,Ge\kern -0.1em V\!/}c}}\xspace}
\newcommand{\mevc}{\ensuremath{{\mathrm{\,Me\kern -0.1em V\!/}c}}\xspace}
\newcommand{\gevcc}{\ensuremath{{\mathrm{\,Ge\kern -0.1em V\!/}c^2}}\xspace}
\newcommand{\gevgevcccc}{\ensuremath{{\mathrm{\,Ge\kern -0.1em V^2\!/}c^4}}\xspace}
\newcommand{\mevcc}{\ensuremath{{\mathrm{\,Me\kern -0.1em V\!/}c^2}}\xspace}

\def\mum  {\ensuremath{{\,\upmu\mathrm{m}}}\xspace}

\def\mbarn{\ensuremath{\mathrm{ \,mb}}\xspace}

\def\invnb {\ensuremath{\mbox{\,nb}^{-1}}\xspace}

\newcommand{\chisqip}{\ensuremath{\chi^2_{\text{\ensuremath{\rm IP}}}}\xspace}

\def\deriv {\ensuremath{\mathrm{d}}}

\def\gsim{{~\raise.15em\hbox{$>$}\kern-.85em
          \lower.35em\hbox{$\sim$}~}\xspace}
\def\lsim{{~\raise.15em\hbox{$<$}\kern-.85em
          \lower.35em\hbox{$\sim$}~}\xspace}

\def\ptot       {\mbox{$p$}\xspace}
\def\pt         {\mbox{$p_{\mathrm{ T}}$}\xspace}

\newcommand{\lum} {\ensuremath{\mathcal{L}}\xspace}

\def\evtgen     {\mbox{\textsc{EvtGen}}\xspace}

\def\geant      {\mbox{\textsc{Geant4}}\xspace}

\def\photos     {\mbox{\textsc{Photos}}\xspace}

\def\pythia     {\mbox{\textsc{Pythia}}\xspace}

\def\tell1  {TELL1\xspace}
\def\ukl1   {UKL1\xspace}

\newcommand{\ie}{\mbox{\itshape i.e.}\xspace}

\newcommand{\DtoKPiAll}{{\ensuremath{\Dz\to K^\mp\pi^\pm}}\xspace}
\newcommand{\DtoKPi}{{\ensuremath{\Dz\to K^+\pi^-}}\xspace}

\newcommand{\DtoKPiP}{{\ensuremath{\Dz\to K^-\pi^+}}\xspace}
\newcommand{\pPb}{{\ensuremath{p\mathrm{Pb}}}\xspace}
\newcommand{\Tev}{\ensuremath{\mathrm{\,Te\kern -0.1em V}}\xspace}

\newcommand{\DfromB}{$\Dz$-from-$b$\xspace}
\newcommand{\IPDz}{{\ensuremath{\log_{10}(\chisqip(\Dz))}}\xspace}
\newcommand{\snn}{{\ensuremath{\sqrt{s_{\mathrm{NN}}}}}\xspace}

\def\be{\begin{equation}}
\def\ee{\end{equation}}

\def\bi{\begin{itemize}}
\def\ei{\end{itemize}}
\def\bc{\begin{center}}
\def\ec{\end{center}}
\def\and{\/\mbox{and}}

\newcommand{\squishlist}{
 \begin{list}{$\bullet$}
  { \setlength{\itemsep}{0pt}
     \setlength{\parsep}{3pt}
     \setlength{\topsep}{3pt}
     \setlength{\partopsep}{0pt}
     \setlength{\leftmargin}{1.5em}
     \setlength{\labelwidth}{1em}
     \setlength{\labelsep}{0.5em} } }

\newcommand{\squishlisttwo}{
 \begin{list}{$\bullet$}
  { \setlength{\itemsep}{0pt}
     \setlength{\parsep}{0pt}
    \setlength{\topsep}{0pt}
    \setlength{\partopsep}{0pt}
    \setlength{\leftmargin}{2em}
    \setlength{\labelwidth}{1.5em}
    \setlength{\labelsep}{0.5em} } }

\newcommand{\squishend}{
  \end{list}  }

\newcommand{\TeV}{\ensuremath{\mathrm{Te\kern -0.1em V}}}
\newcommand{\GeV}{\ensuremath{\mathrm{Ge\kern -0.1em V}}}
\newcommand{\MeV}{\ensuremath{\mathrm{Me\kern -0.1em V}}}

\newcommand{\met}{\mbox{${\hbox{$E$\kern-0.6em\lower-.1ex\hbox{/}}}_{\rm T}\:$}}
\newcommand{\metx}{\mbox{${\hbox{$E$\kern-0.6em\lower-.1ex\hbox{/}}}_T\:$}}
\newcommand{\mety}{\mbox{${\hbox{$E$\kern-0.6em\lower-.1ex\hbox{/}}}_T\:$}}

\newcommand{\bit}{\begin{itemize}}
\newcommand{\eit}{\end{itemize}}
\newcommand{\bce}{\begin{center}}
\newcommand{\beqn}{\begin{eqnarray*}}
\newcommand{\eeqn}{\end{eqnarray*}}
\newcommand{\beq}{\begin{equation}}
\newcommand{\ece}{\end{center}}

\usepackage{cite} 
\usepackage{mciteplus}

\usepackage{longtable} 
\usepackage{amsmath}
\usepackage{varwidth}

\begin{document}

\renewcommand{\thefootnote}{\fnsymbol{footnote}}
\setcounter{footnote}{1}


\begin{titlepage}
\pagenumbering{roman}

\vspace*{-1.5cm}
\centerline{\large EUROPEAN ORGANIZATION FOR NUCLEAR RESEARCH (CERN)}
\vspace*{1.5cm}
\noindent
\begin{tabular*}{\linewidth}{lc@{\extracolsep{\fill}}r@{\extracolsep{0pt}}}
\ifthenelse{\boolean{pdflatex}}
{\vspace*{-2.7cm}\mbox{\!\!\!\includegraphics[width=.14\textwidth]{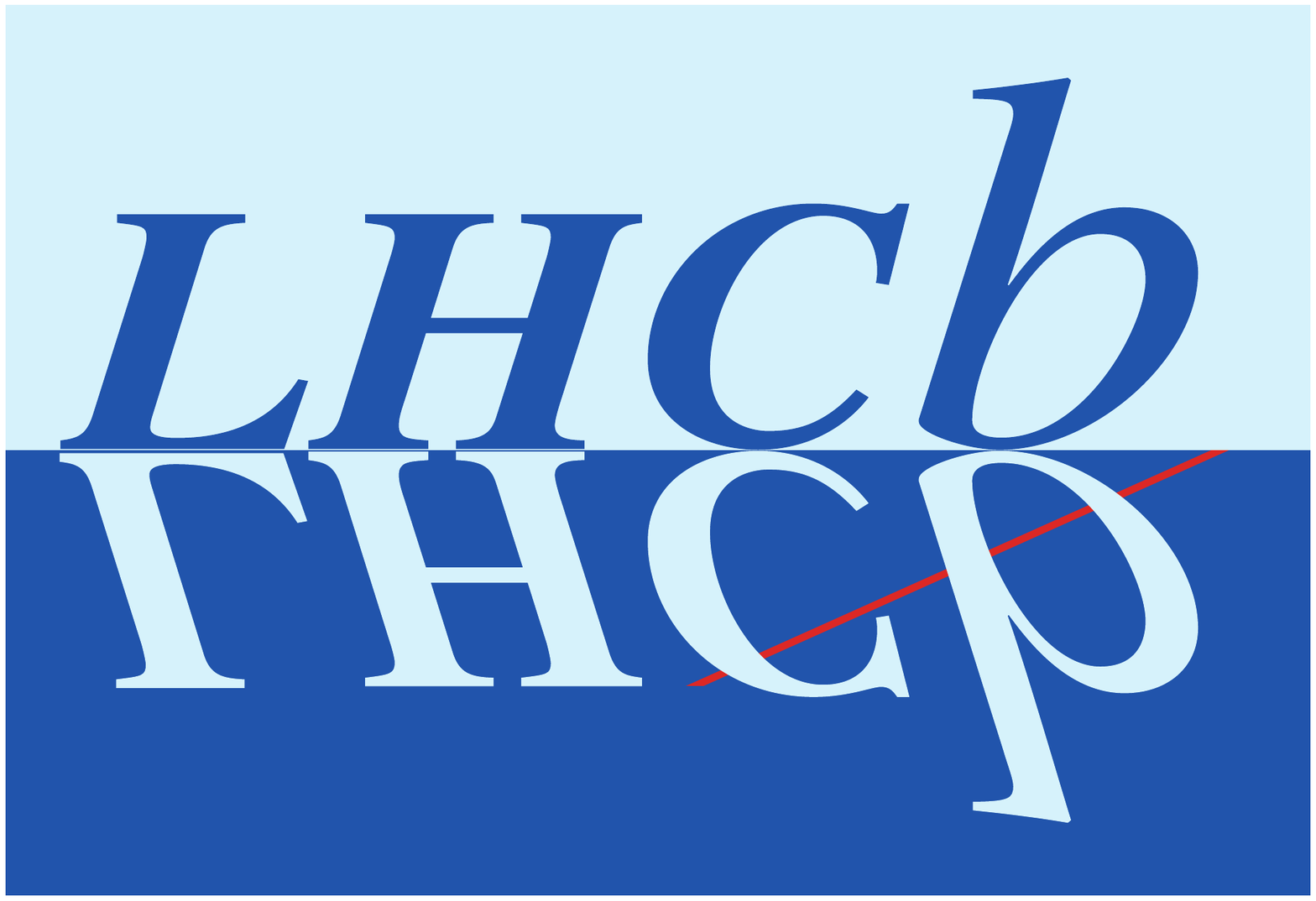}} & &}%
{\vspace*{-1.2cm}\mbox{\!\!\!\includegraphics[width=.12\textwidth]{lhcb-logo.eps}} & &}%
\\
 & & CERN-EP-2017-135 \\  
 & & LHCb-PAPER-2017-015 \\  
 & & 10$^{th}$ July 2017
\end{tabular*}

\vspace*{3.0cm}

{\normalfont\bfseries\boldmath\huge
\begin{center}
 Study of prompt $\Dz$ meson production in $p\mathrm{Pb}$ collisions at $\snn=5\,\tev$
\end{center}
}

\vspace*{2.0cm}

\begin{center}
The LHCb collaboration\footnote{Authors are listed at the end of this paper.}
\end{center}

\vspace{\fill}

\begin{abstract}
  \noindent
  Production of prompt $\Dz$ mesons is studied in proton-lead 
and lead-proton
  collisions recorded at the LHCb detector at the LHC. 
  The data sample corresponds to an integrated luminosity
  of $1.58\pm0.02\invnb$ recorded at a 
  nucleon-nucleon centre-of-mass energy of $\snn=5\tev$.
  Measurements of the differential cross-section, the 
  forward-backward production ratio and the nuclear modification factor
 are reported  using $\Dz$ 
  candidates with transverse momenta less than $10\gevc$ 
  and rapidities in the ranges $1.5<y^*<4.0$ and $-5.0<y^*<-2.5$
  in the nucleon-nucleon centre-of-mass
  system. 
  
\end{abstract}

\vspace*{2.0cm}

\begin{center}
  Submitted to JHEP
\end{center}

\vspace{\fill}

{\footnotesize 
\centerline{\copyright~CERN on behalf of the \lhcb collaboration, licence \href{http://creativecommons.org/licenses/by/4.0/}{CC-BY-4.0}.}}
\vspace*{2mm}

\end{titlepage}


\newpage
\setcounter{page}{2}
\mbox{~}
%
\cleardoublepage


\renewcommand{\thefootnote}{\arabic{footnote}}
\setcounter{footnote}{0}



\pagestyle{plain} 
\setcounter{page}{1}
\pagenumbering{arabic}

\section{Introduction}
\label{sec:Introduction}

Charm hadrons produced in hadronic and nuclear collisions
are excellent probes 
to study nuclear matter in extreme conditions. 
The differential cross-sections of $c$-quark production 
in $pp$ or $p\bar{p}$ collisions
have been calculated
based on perturbative quantum chromodynamics (QCD)
 and collinear or $k_\mathrm{T}$ factorisation~\cite{Kniehl:2005ej,
Kniehl:2012ti,Cacciari:1998it,Cacciari:2003zu,Cacciari:2012ny,Maciula:2013wg}.
These phenomenological models~\cite{Andronic:2015wma}
are also able to predict the differential cross-section of $c$-quark 
production including most of the commonly assumed
``cold nuclear matter'' (CNM) effects in nuclear collisions,
where CNM effects related to the parton
flux differences and other effects come into play.  
Since heavy quarks are produced 
at a time scale of approximately  0.1~fm/$c$ after the 
collision, they are ideal to examine 
hot nuclear matter, the so-called ``quark-gluon 
plasma'' (QGP), by studying how they traverse 
this medium and interact with it right after 
their formation.
These studies require a thorough understanding of
the CNM effects, which can be investigated in systems
where the formation of QGP is not expected.
In addition, a precise quantification of CNM effects 
would significantly improve 
the understanding of charmonium 
and open-charm production by 
confirming or discarding the possibility that  
the suppression pattern in the production of quarkonium 
states, like \jpsi, at the SPS, RHIC and LHC is due to 
QGP formation~\cite{Andronic:2015wma}.

The study of CNM effects is best performed in collisions of protons 
with heavy nuclei like lead, where 
the most studied CNM effects,
such as gluon saturation~\cite{Kharzeev:2005zr,Fujii:2006ab} 
and in-medium energy         
loss~\cite{Arleo:2012rs}
in initial- and final-state radiation~\cite{Gavin:1991qk,Vogt:1999dw},
are more evident.
Phenomenologically, collinear parton distributions are often used
to describe the nuclear modification of the parton flux in the nucleus. 
The modification with respect to the free nucleon depends on
the parton fractional longitudinal momentum, $x$, and the atomic mass number
of the nucleus $A$~\cite{Armesto:2006ph,Malace:2014uea}.
In the low-$x$ region, down to $x\approx  10^{-5}-10^{-6}$, which 
is accessible at LHC energies, stronger 
onset of gluon saturation~\cite{Fujii:2013yja,Tribedy:2011aa,Albacete:2012xq,Rezaeian:2012ye}
is expected to play a
major role. Its effect can be quantified by studying production 
of $\Dz$ mesons at low transverse momentum \pt~\cite{ALICE:2012ab}, ideally down to zero $\pt$.
The in-medium energy loss occurs when the partons lose energy in the cold medium through both
initial- and final-state radiation.

CNM effects have been investigated in detail at the 
RHIC collider in $pp$ and $d$Au 
collisions~\cite{Andronic:2015wma,Averbeck:2013oga} at a nucleon-nucleon centre-of-mass 
energy of $\snn=200$\gev.
Most recently, CNM effects were measured in 
$p$Pb collisions at the LHC for quarkonium and heavy flavour production~\cite{Adamova:2017uhu,Adam:2016ich,
Adam:2016ohd,Adam:2016mkz, Adam:2015jsa, Adam:2015iga,Abelev:2014oea, Abelev:2014hha,Abelev:2013yxa, Abelev:2014zpa, Aad:2015ddl,
Chatrchyan:2013nza,Sirunyan:2017mzd, Sirunyan:2016fcs, Khachatryan:2015sva,Khachatryan:2015uja}.
The ALICE experiment~\cite{Abelev:2014hha}
  studied $D$ meson production in 
$p$Pb collisions at $\snn=5$\tev 
in the region $−0.96<y^*<0.04$
for $\pt>2\gevc$, where $y^*$ is the rapidity
of the $D$ meson  
defined in the centre-of-mass system of the 
colliding nucleons.
Their results suggest that the 
suppression observed in PbPb collisions is due to 
hot nuclear matter effects, \ie QGP formation.
Results on leptons from semileptonic heavy-flavour decays at
various rapidities 
are also available~\cite{Adam:2015qda,Adam:2016wyz,Acharya:2017hdv}.

In this paper the measurement of the cross-section 
and of the nuclear modification factors of ``prompt'' $D^0$
mesons, \ie  those  directly produced  in proton-lead  collisions 
and not coming from decays of $b$-hadrons, is presented. The measurement
is performed at $\snn=5$~TeV
with the LHCb~\cite{Alves:2008zz} detector at the LHC.
Depending on the direction of the proton and $^{208}$Pb beams and due to the different
energies per nucleon in the two beams, the LHCb detector covers two different
acceptance regions in the nucleon-nucleon rest frame,
\newpage
\begin{itemize}
\item  $1.5 < y^{\ast} < 4.0$, denoted as ``forward'' beam configuration,
\item $- 5.0 < y^{\ast} < - 2.5$, denoted as  ``backward'' beam configuration,
\end{itemize}
where the rapidity $y^\ast$ is defined with respect to the direction of the proton beam,
\noindent The measurement is performed in the
range of \Dz transverse momentum $\pt<10\gevc$,
in both backward and forward collisions.

\section{Detector and data samples}
\label{sec:Detector}
The \lhcb detector~\cite{Alves:2008zz,LHCb-DP-2014-002} is a single-arm forward
spectrometer covering the \mbox{pseudorapidity} range $2<\eta <5$,
designed for the study of particles containing \bquark or \cquark
quarks. The detector includes a high-precision tracking system
consisting of a silicon-strip vertex detector surrounding the $pp$
interaction region (VELO), a large-area silicon-strip detector (TT) located
upstream of a dipole magnet with a bending power of about
$4{\mathrm{\,Tm}}$, and three stations of silicon-strip detectors and straw
drift tubes (OT) placed downstream of the magnet.
The tracking system provides a measurement of momentum, \ptot, of charged particles with
a relative uncertainty that varies from 0.5\% at low momentum to 1.0\% at 200\gevc.
The minimum distance of a track to a primary vertex (PV), the impact parameter, is 
measured with a resolution of $(15+29/\pt)\mum$,
where \pt is the component of the momentum transverse to the beam, in\,\gevc.
Different types of charged hadrons are distinguished using information
from two ring-imaging Cherenkov detectors. 
Photons, electrons and hadrons are identified by a calorimeter system consisting of
scintillating-pad and preshower detectors, an electromagnetic
calorimeter and a hadronic calorimeter. Muons are identified by a
system composed of alternating layers of iron and multiwire
proportional chambers.
The online event selection is performed by a trigger~\cite{LHCb-DP-2012-004}, 
which consists of a hardware stage, based on information from the calorimeter and muon
systems, followed by a software stage, which applies a full event
reconstruction.

The data sample used in this analysis consists of $p$Pb
collisions collected in early 2013, corresponding to integrated luminosities of 
  ($1.06\pm 0.02$)\invnb and ($0.52\pm0.01$)\invnb for 
the forward and backward colliding beam configurations,
respectively.
The luminosity has been determined using the same method as in 
the LHCb measurement of \jpsi production in $p$Pb collisions~\cite{LHCb-PAPER-2013-052},
with a precision of about 2\%. 
The instantaneous luminosity during the period of 
data taking was
around 5$\times 10^{27}$~cm$^{-2}$~s$^{-1}$, 
which led to an event rate that was three orders of magnitude
 lower than in nominal \lhcb $pp$ operation.
Therefore, the hardware trigger simply rejected
empty events,
while the next level software trigger accepted all events with at
least one track in the VELO.

For the analyses presented below, simulated samples of $pp$
collisions at 8 TeV are used to determine geometrical acceptance
and reconstruction efficiencies.  
Effects due to the different track multiplicity 
distributions in the $pp$ and $p$Pb
collision data and the effects of the asymmetric beam energies in $p$Pb
collisions are taken into account as described later.
In the simulation, $pp$ collisions are generated using 
\pythia~\cite{Sjostrand:2006za,*Sjostrand:2007gs} 
with a specific \lhcb configuration~\cite{LHCb-PROC-2010-056}.  
Decays of hadronic particles are described by \evtgen~\cite{Lange:2001uf}, 
in which final-state radiation is generated using 
\photos~\cite{Golonka:2005pn}.
The interaction of the generated particles with the detector, and 
its response, are implemented using the \geant toolkit~\cite{Allison:2006ve, *Agostinelli:2002hh,LHCb-PROC-2011-006}.

\section{Cross-section determination}\label{sec:xsec}

The double-differential cross-section for prompt $\Dz$ production in a given $(\pt,y^{*})$ kinematic bin
is defined as 

\begin{equation}\label{eq:cross-section}
  \frac{\deriv^2\sigma}{\deriv \pt\deriv y^*} 
  = \frac{N(\DtoKPiAll)}
  {\lum\times\etot\times\BR(\DtoKPiAll)\times\Delta \pt\times\Delta y^*},
\end{equation}
where 
$N(\DtoKPiAll)$ is the number of prompt \Dz signal candidates reconstructed through the
 $\DtoKPiAll$ decay channels\footnote{Charge conjugation is 
implied throughout this document if not otherwise specified.},
$\etot$ is the total \Dz detection efficiency,
$\lum$ is the integrated luminosity,
$\BR(\DtoKPiAll)=(3.94\pm0.04)\%$ is the sum of the branching 
fractions of the decays $\DtoKPiP$ and $\DtoKPi$~\cite{PDG2016},
$\Delta\pt=1\gevc$ is the bin width of the \Dz transverse momentum,
and $\Delta y^*=0.5$ is the bin width of the \Dz rapidity.
The  rapidity $y^*$ is defined in the nucleon-nucleon centre-of-mass frame, where the positive direction is that 
of the proton beam.
The measurement is performed in the \Dz kinematic region defined by 
 $\pt<10\gevc$ and rapidities  $1.5<y^*<4.0$ for the forward sample 
and $-5.0<y^*<-2.5$ for 
the backward sample.

The total cross-section over a specific kinematic range is determined by 
integration of the double-differential cross-section. The nuclear 
modification factor, $R_{p{\rm Pb}}$, is the ratio of the \Dz production cross-section
in forward or backward collisions to that in $pp$ at the 
same nucleon-nucleon centre-of-mass energy $\snn$
\begin{equation}
R_{p{\rm Pb}} (\pt,y^*) \equiv \frac{1}{A} \frac{{\rm d}^2 \sigma_{p{\rm Pb}}(\pt,y^*)/{\rm d}\pt{\rm d}y^*}
{{\rm d}^2\sigma_{pp}(\pt,y^*)/{\rm d}\pt{\rm d}y^*},
\end{equation}
where $A$=208 is the atomic mass number of the lead nucleus.
The forward-backward production ratio, $R_{\rm FB}$, 
is defined as
\begin{equation}
R_{\rm FB} (\pt,y^*) \equiv \frac{{\rm d}^2 \sigma_{p{\rm Pb}}(\pt,+|y^*|)/{\rm d}\pt{\rm d}y^*}{{\rm d}^2 \sigma_{{\rm Pb}p}(\pt,-|y^*|)/{\rm d}\pt{\rm d}y^*},
\end{equation}
where $\sigma_{p{\rm Pb}}$ and $\sigma_{{\rm Pb}p}$ indicate the 
cross-sections in the forward and backward configurations respectively, 
measured in a common rapidity range. 
The \Dz candidates are selected according to the same requirements
as used in the $\Dz$ production cross-section measurements 
in $pp$ collisions at $\sqrt{s}=7\tev$ \cite{LHCb-PAPER-2012-041} and
$\sqrt{s}=13\tev$\cite{LHCb-PAPER-2015-041}.
The kaon and pion tracks from the \Dz candidate and the vertex they form are
both required to be of good quality. The requirements set on 
particle identification (PID) criteria are tighter
than in $pp$ collisions
 to increase the signal-over-background ratio 
given the high detector occupancy observed 
in $p$Pb collisions. 
\begin{figure}[!tbp]
\centering
\begin{minipage}[t]{0.99\textwidth}
\centering
\begin{tabular}{cc}
\includegraphics[width=0.5\textwidth]{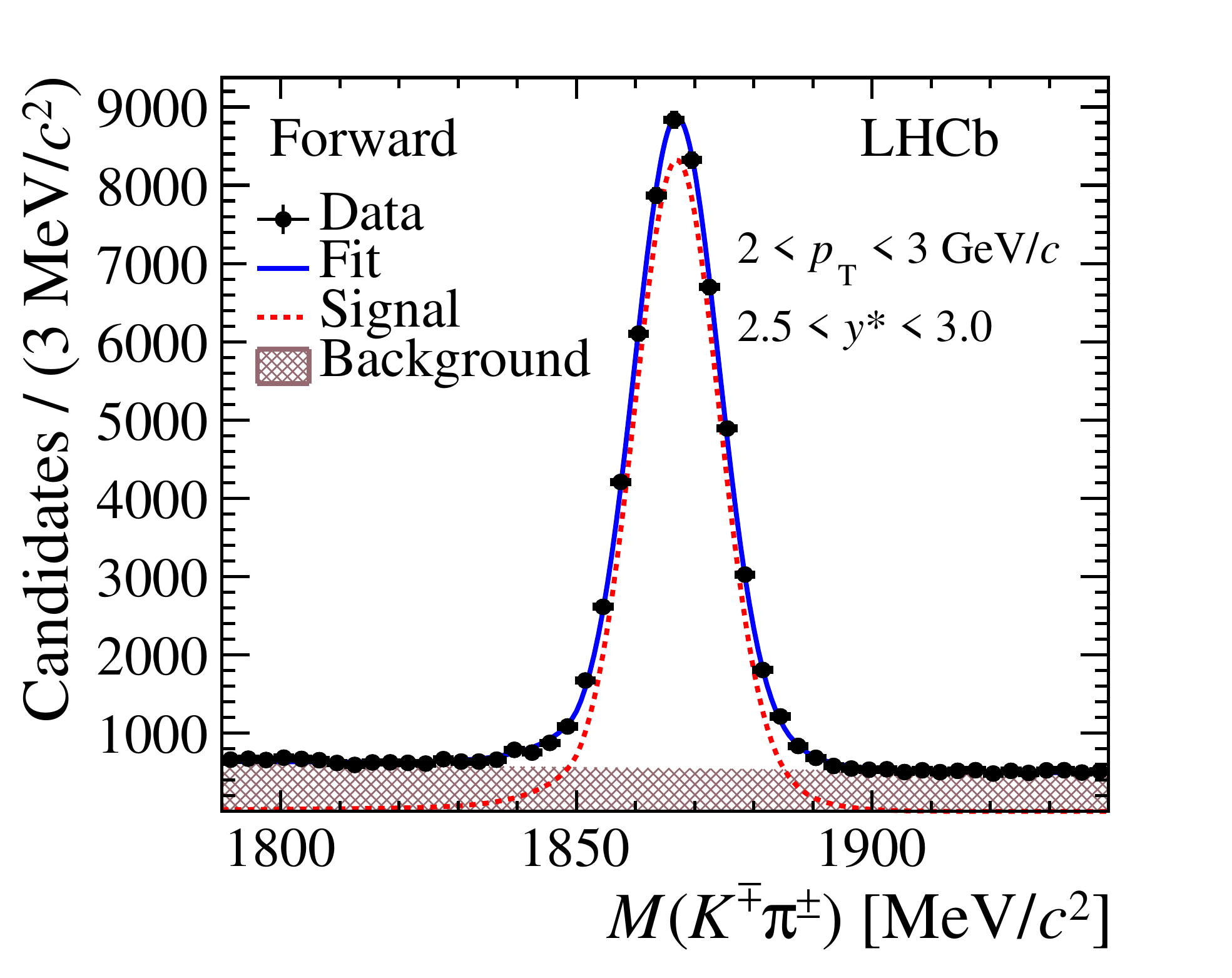}&
\includegraphics[width=0.5\textwidth]{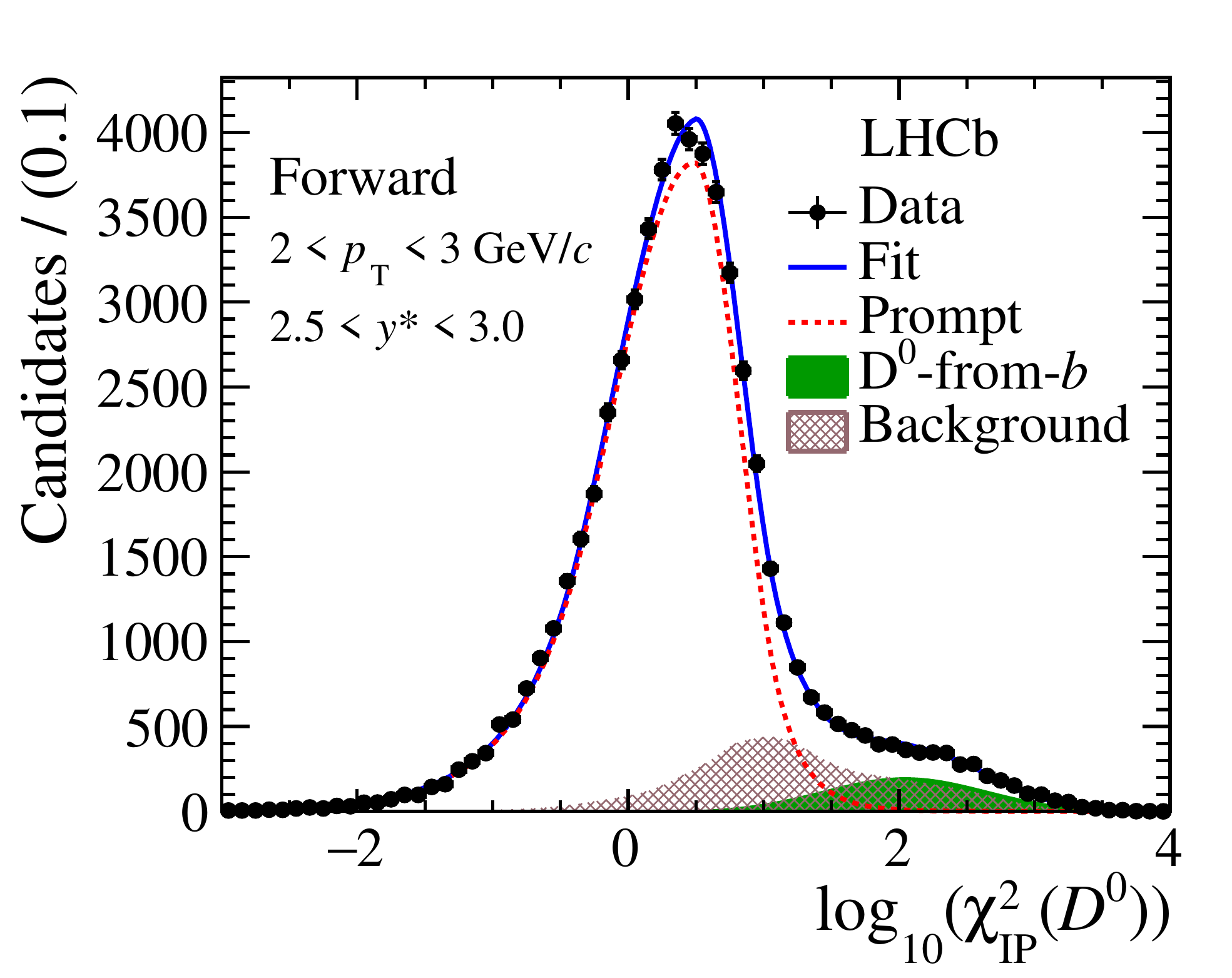}\\
\end{tabular}
\end{minipage}
\caption{The (left)  $M(K^\mp\pi^\pm)$ and (right) $\IPDz$ distributions and the 
fit result for the inclusive $\Dz$ mesons in the forward data sample in the kinematic range of $2<\pt<3\gevc$
and $2.5<y^*<3.0$.}
\label{fig:MassIPFittPA}
\end{figure}
\begin{figure}[!tbp]
\centering
\begin{minipage}[t]{0.99\textwidth}
\centering
\begin{tabular}{cc}
\includegraphics[width=0.5\textwidth]{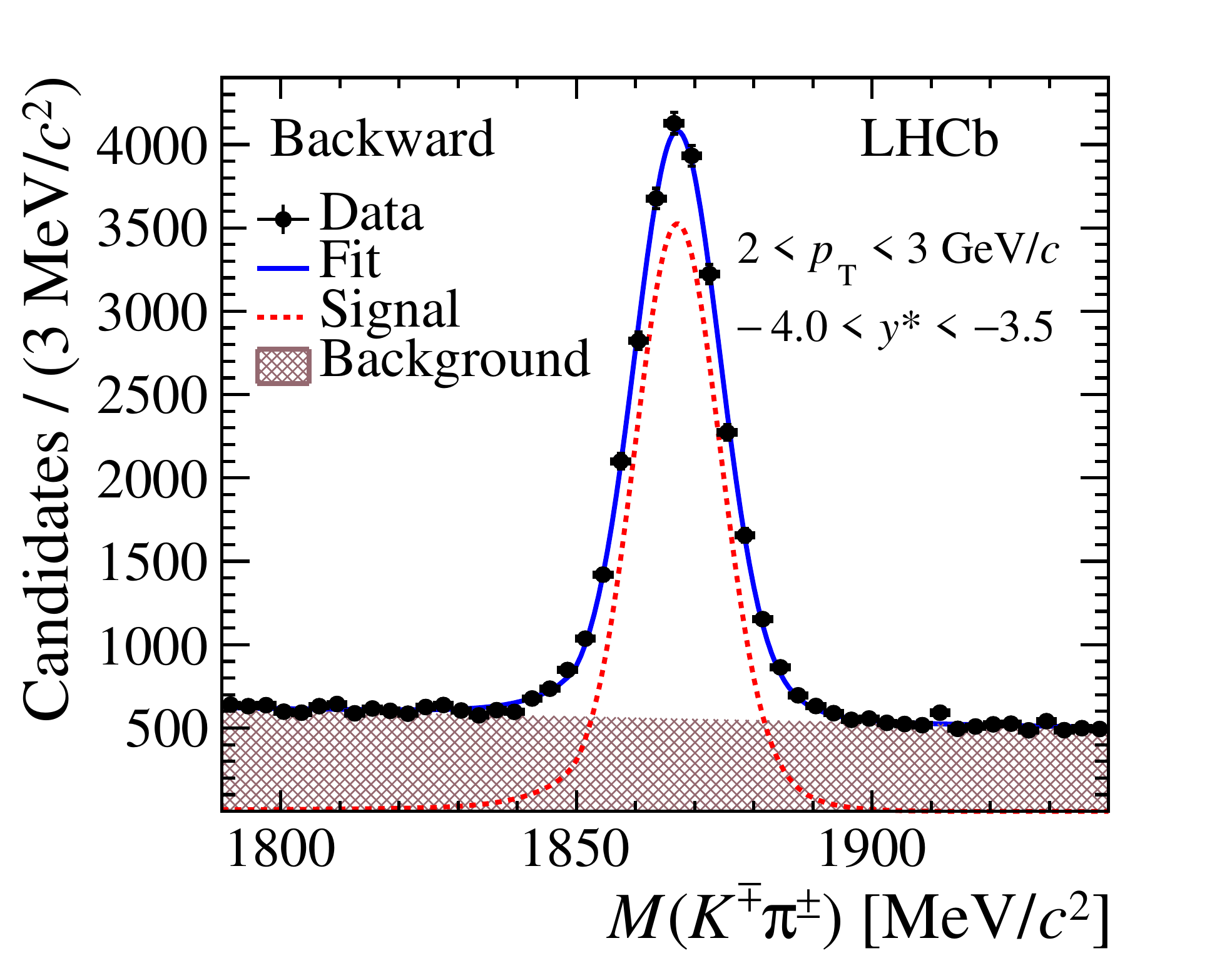}&
\includegraphics[width=0.5\textwidth]{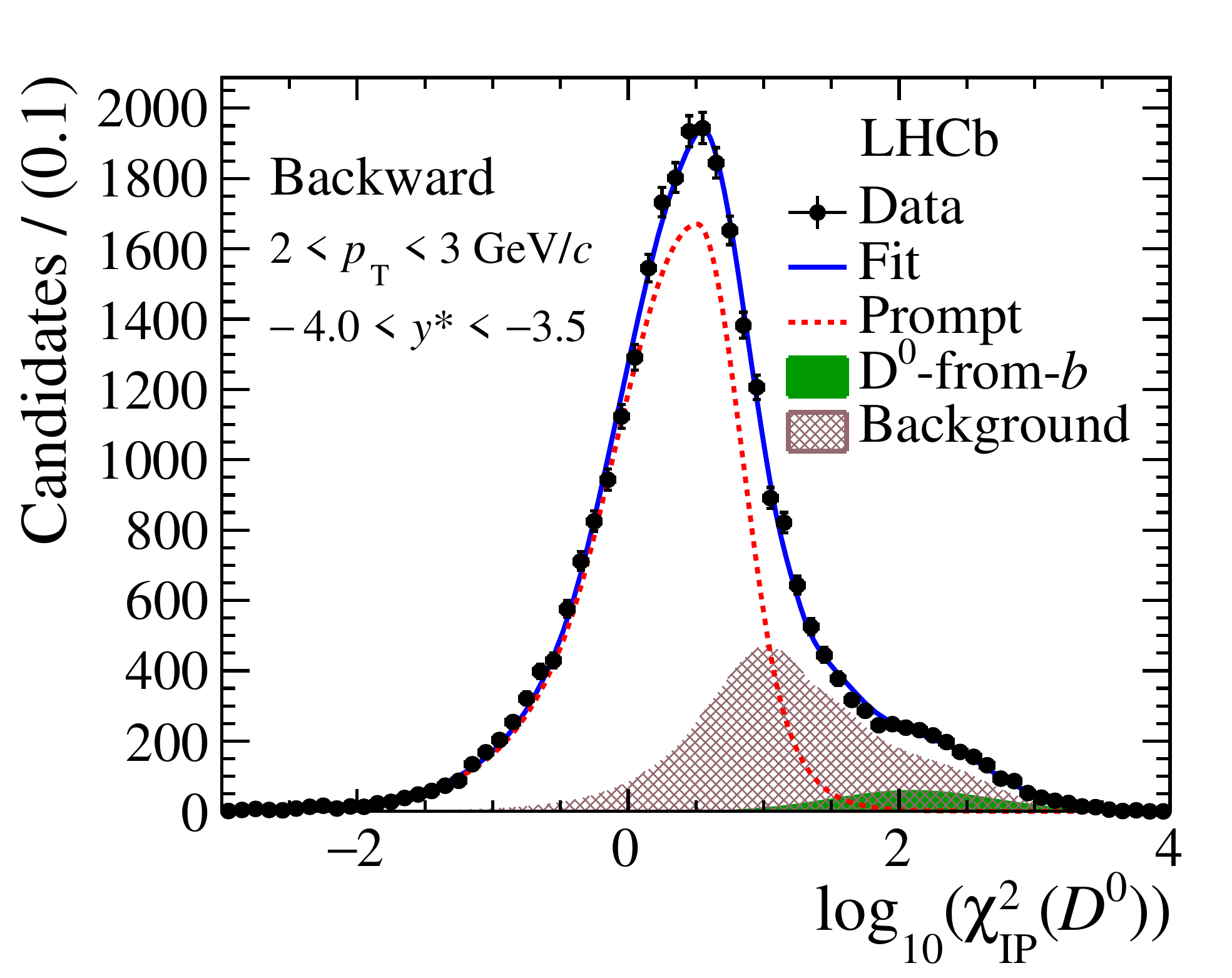}\\
\end{tabular}
\end{minipage}
\caption{The (left)  $M(K^\mp\pi^\pm)$ and (right) $\IPDz$ distributions and the 
fit result for the inclusive $\Dz$ mesons in the backward data sample in the kinematic range of $2<\pt<3\gevc$
and $-4.0<y^*<-3.5$.}
\label{fig:MassIPFittAP}
\end{figure}

\newcommand{\rhoL}{\ensuremath{\rho_\mathrm{L}}\xspace}
\newcommand{\rhoR}{\ensuremath{\rho_\mathrm{R}}\xspace}

The signal yield is
determined from an extended unbinned maximum likelihood fit to the
distribution of the invariant mass $M(K^\mp\pi^\pm)$. 
The fraction of nonprompt $\Dz$ 
mesons originating from $b$-hadron decays, called 
\DfromB\ in the following, 
is determined from the $\IPDz$ distribution, where $\chisqip(D^0)$ is 
defined as 
the difference in vertex-fit $\chi^2$ of a given PV
computed with and without the \Dz meson candidate\cite{LHCb-PAPER-2012-041,LHCb-PAPER-2015-041}. 
On average, prompt \Dz mesons have much 
smaller  $\chisqip(D^0)$ values than \DfromB.
The fit is performed in two steps. First, the invariant mass distributions 
are fitted to determine
the $\Dz$ meson inclusive yield and the number of  background candidates, 
then the $\IPDz$ 
fit is performed for candidates with mass within $\pm20\mevcc$ around
 the fitted value of the \Dz mass.
In the $\IPDz$ fit, 
the number of background candidates is constrained to the value obtained 
from the
invariant mass fit, scaled to the selected mass range.

The distribution of $\IPDz$ is shown in the right-hand plots of Figs.~\ref{fig:MassIPFittPA} and \ref{fig:MassIPFittAP}
for the forward and backward samples, respectively. 
 The 
signal shape in the $M(K^\mp\pi^\pm)$ distributions is described by a Crystal Ball (CB) 
function~\cite{Skwarnicki:1986xj} plus a Gaussian. The mean 
is the same 
for both functions, and the ratios of widths and 
tail parameters are fixed 
following simulation studies, as in 
previous LHCb analyses~\cite{LHCb-PAPER-2012-041,LHCb-PAPER-2015-041}.
The width, mean, and signal yields are left free to vary.
The background is described by a linear function. The candidates are 
fitted in the range 1792--1942\mevcc.
The invariant mass distributions in the inclusive forward and backward samples are 
shown in the left-hand plots of Figs.~\ref{fig:MassIPFittPA} and \ref{fig:MassIPFittAP}
respectively.

The fits to the invariant mass and $\IPDz$ distributions 
are performed independently
in each bin of $(\pt,y^*)$ of the $\Dz$ meson. The contribution of 
the \DfromB\ component increases with transverse
momentum up to 10\%.
The $\IPDz$ shapes for the prompt \Dz meson signal candidates 
are estimated using 
the simulation and modelled with a modified Gaussian function
\begin{equation}
f_\mathrm{}(x; \mu, \sigma, \epsilon, \rhoL, \rhoR) =
  \begin{cases}
    e^{\frac{\rhoL^{2}}{2} + \rhoL\frac{x - \mu}{(1 - \epsilon)\sigma}} & x < \mu - (\rhoL\sigma(1 - \epsilon)), \\
    e^{-\left(\frac{x - \mu}{\sqrt{2}\sigma(1 - \epsilon)}\right)^{2}} & \mu - (\rhoL\sigma(1 - \epsilon)) \leq x < \mu, \\
    e^{-\left(\frac{x - \mu}{\sqrt{2}\sigma(1 + \epsilon)}\right)^{2}} & \mu \leq x < \mu + (\rhoR\sigma(1 + \epsilon)), \\
    e^{\frac{\rhoR^{2}}{2} - \rhoR\frac{x - \mu}{(1 + \epsilon)\sigma}} & x \geq \mu + (\rhoR\sigma(1 + \epsilon)),
  \end{cases}
\end{equation}
where the values of $\epsilon$, \rhoL and \rhoR are fixed to the values obtained
in the simulation and $\mu$ and $\sigma$ are free parameters.
The $\IPDz$ distribution for the \DfromB\ component is 
described by a Gaussian function. 
The shape of the combinatorial background is estimated using the distribution of
candidates
with mass in the ranges
1797--1827\mevcc and 1907--1937\mevcc, \ie between 40 and 70~\mevcc 
away from the observed $\Dz$ meson mass. 

The total efficiency $\etot$ in Eq.~\ref{eq:cross-section} includes the 
effects of geometrical
acceptance and the efficiencies of the trigger,  of the reconstruction
and of the PID criteria used in the analysis.
The analysis uses a minimum activity 
trigger, whose efficiency for events containing a \Dz meson is found to be 
100\%.
The geometrical acceptance and
reconstruction efficiencies are estimated using $pp$ simulated samples, validated with data. 
 The difference between the distributions of the track multiplicity in the $p$Pb and $pp$
collisions is accounted for by studying the efficiency
in bins of the track multiplicity, and weighting the efficiency according to
the  multiplicity distributions seen in $p$Pb
and Pb$p$ data.
The related systematic uncertainties are discussed in 
Sec.~\ref{sec:syst}.
The PID efficiency is estimated 
using a calibration sample of \Dz meson decays selected in data without 
PID requirements~\cite{LHCb-DP-2014-002},
and collected in the 
same period as the \pPb sample used for the analysis.
The PID selection efficiency is calculated by using the 
$K^\mp$ and $\pi^\pm$ single-track efficiencies from calibration data, 
and averaging them according to the kinematic 
distributions observed in the simulation in each 
$\Dz$ $(\pt,y^{*})$ bin.

\section{Systematic Uncertainties}\label{sec:syst}

The systematic uncertainties affecting the cross-sections 
 are listed 
in Table~\ref{tab:SystematicSummary}.
\begin{table}[tbh]
\caption{Summary of systematic and statistical uncertainties on the cross-section. 
The ranges indicate the variations between bins, with the uncertainty on average 
increasing with rapidity and momentum.}
\centering
\begin{tabular}{l|cc}
\hline
Source & \multicolumn{2}{c}{Relative uncertainty (\%)}\\
\hline
    & Forward    & Backward \\
{\it Correlated between bins}    &  &  \\
Invariant mass fits                  & 0.0 $-$\phantom{0}5.0   & 0.0 $-$\phantom{0}5.0\\
\IPDz fits            & 0.0 $-$\phantom{0}5.0   & 0.0 $-$\phantom{0}5.0\\
Tracking efficiency                  & 3.0   & 5.0\\
PID efficiency             &\, 0.6 $-$ 17.0   &\,0.6 $-$ 30.0 \\
Luminosity                 & 1.9   & 2.1\\
$\mathcal{B}(\DtoKPiAll)$     & 1.0     & 1.0\\
{\it Uncorrelated between bins}    &     &  \\
Simulation sample size            & 1.0 $-$\phantom{0}4.0       &1.0 $-$\phantom{0}5.0\\
\hline
Statistical uncertainty    &\,0.5 $-$ 20.0    & \,1.0 $-$ 20.0\\
\hline
\end{tabular}
\label{tab:SystematicSummary}
\end{table}
They are evaluated 
separately for the backward and forward samples unless otherwise 
specified.
The systematic uncertainty associated to the 
determination of the signal yield 
has contributions from the signal and 
background models.
The uncertainty 
associated to the modelling of the signal
is studied by using alternative models of single or sum of two Gaussian
functions to fit the invariant mass
in the forward and backward samples. 
A variation of the parameters which are fixed 
in the default model, within the ranges
indicated by the simulation, is also explored.
The largest difference between the 
nominal and the alternative fits is taken
as the uncertainty on the method, which results in 
a bin-dependent uncertainty, not exceeding 5\%.
The effect due to background modelling in the invariant mass fit is studied by using 
an exponential as an alternative to the linear function.
This uncertainty is found to be negligible.
For the fit to the $\IPDz$ distribution, the \rhoL and \rhoR parameters of the prompt signal component are varied within
the ranges studied in simulation. The distribution of combinatorial backgrounds is studied with candidates in different
background mass regions. The shape of the distribution for the \DfromB\  component is fixed when studying the variation
of its fraction.
The same procedure is followed to estimate the 
uncertainty on the $\IPDz$ fits. 
The systematic uncertainty on the prompt signal yields,
determined by the $\IPDz$ fit, depends on the kinematic bin and is estimated
to be less than 5\% in all cases.

The systematic uncertainty associated with the tracking efficiency has
the components described in the following.  
The efficiency measurement is affected by the
imperfect modelling of the tracking efficiency by simulation, which is corrected using a data-driven method~\cite{LHCb-DP-2013-002},
and the
uncertainty of the correction is 
propagated into an uncertainty on the \Dz yield.
The limited sizes of the simulated samples affect the precision of the efficiency, especially
 in the high multiplicity region. 
Another source of uncertainty is introduced by the choice of variable representing
 the detector occupancy, used to weight the distributions.
The number of tracks and the number of hits in the VELO and in the TT and OT 
 are all considered separately. The largest difference between
the efficiencies when weighted by each of these variables and their 
average, which is the default, is taken as systematic uncertainty.
An additional uncertainty comes from the detector occupancy distribution 
estimated in backward and forward data.
The effects are  summed in quadrature, yielding 
 a total uncertainty on the tracking efficiency of $3\%$ and $5\%$ for the 
forward and backward collision sample respectively.

\label{sec:PIDSys}

The limited size of  the calibration sample, the binning
scheme and the signal fit model used to determine the
$\pi$ and $K$  PID efficiency from the calibration sample,
all contribute to the systematic uncertainty.
The first is evaluated by 
estimating new sets of efficiencies through the
variation of the $\pi$ and $K$ PID efficiencies 
in the calibration sample within the statistical 
uncertainties, the second by using alternative binning
schemes and the third by varying the signal 
function used to determine the signal. 
 The uncertainty is taken to be the quadratic sum of the
three components.
The total PID systematic uncertainty 
ranges between 1\% and 30\% depending on the kinematic 
region and the collision sample.

The relative uncertainty associated with the luminosity measurement is 
approximately $2\%$ for both forward and backward samples.
The relative uncertainty of the branching fraction $\BF(\DtoKPiAll)$ is $1\%$~\cite{PDG2016}. 
The limited size of the simulation sample
introduces uncertainties on the efficiencies which are then propagated to the cross-section measurements; this effect is
negligible for  the central rapidity region but increases in the regions close to the boundaries of 
$\pt$ and $y$, ranging between 1\% and 5\%.

\section{Results}
\label{sec:Result}
\subsection{Production cross-sections}
\begin{figure}[tbp]
\centering
\begin{minipage}[t]{0.49\textwidth}
\centering
\includegraphics[width=1.0\textwidth]{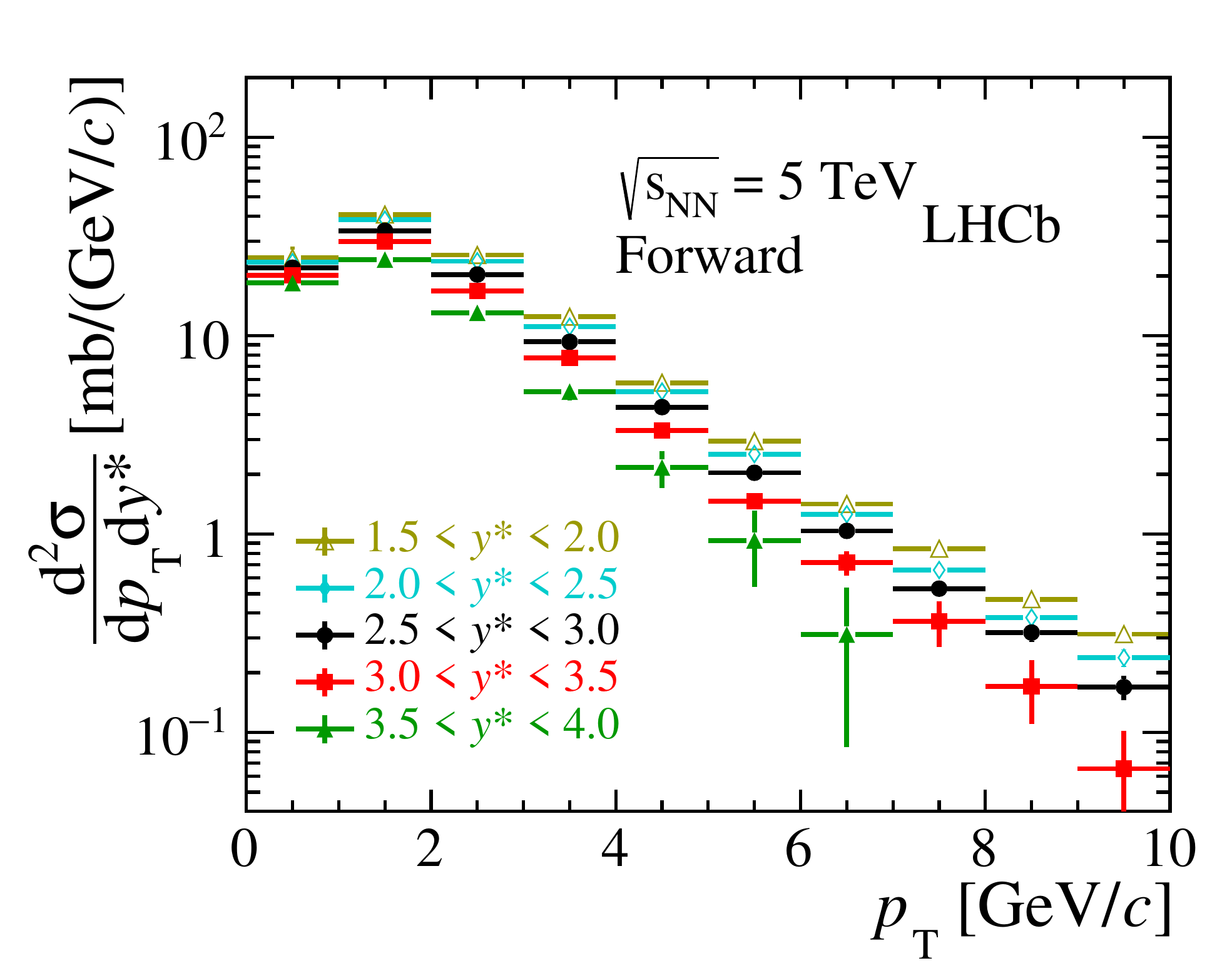}
\end{minipage}
\begin{minipage}[t]{0.49\textwidth}
\centering
\includegraphics[width=1.0\textwidth]{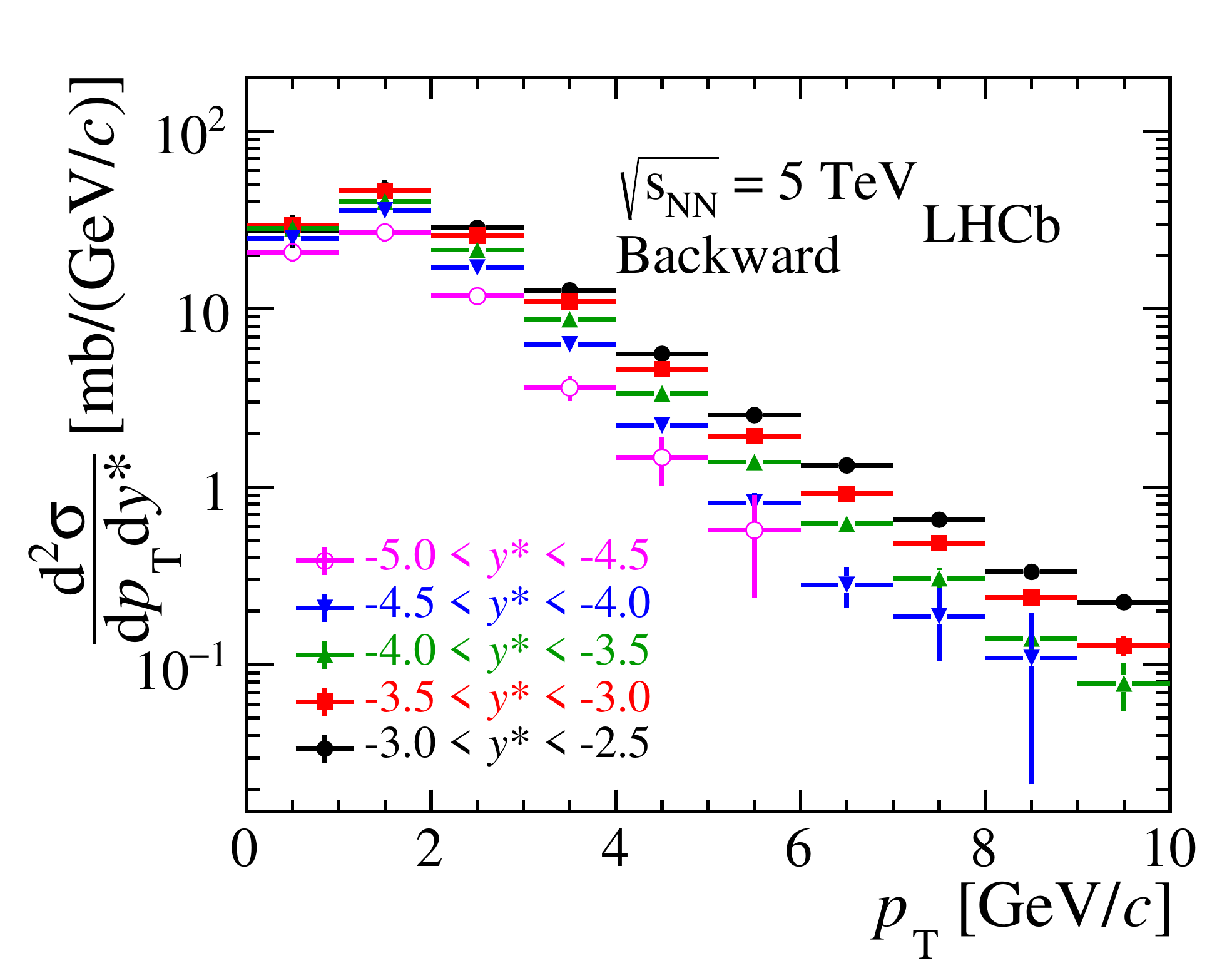}
\end{minipage}
\caption{Double-differential cross-section $\frac{\deriv^2\sigma}{\deriv p_{\mathrm T}\deriv y^*}$\,(mb/(\gevc))  of prompt $\Dz$ meson production in $\pPb$ collisions in the (left) forward and (right) backward
collision samples.
The uncertainty is the quadratic sum of the statistical and systematic components.
}
\label{fig:CrossSection2D}
\end{figure}

\begin{figure}[tbp]
\centering
\begin{minipage}[t]{0.49\textwidth}
\centering
\includegraphics[width=1.0\textwidth]{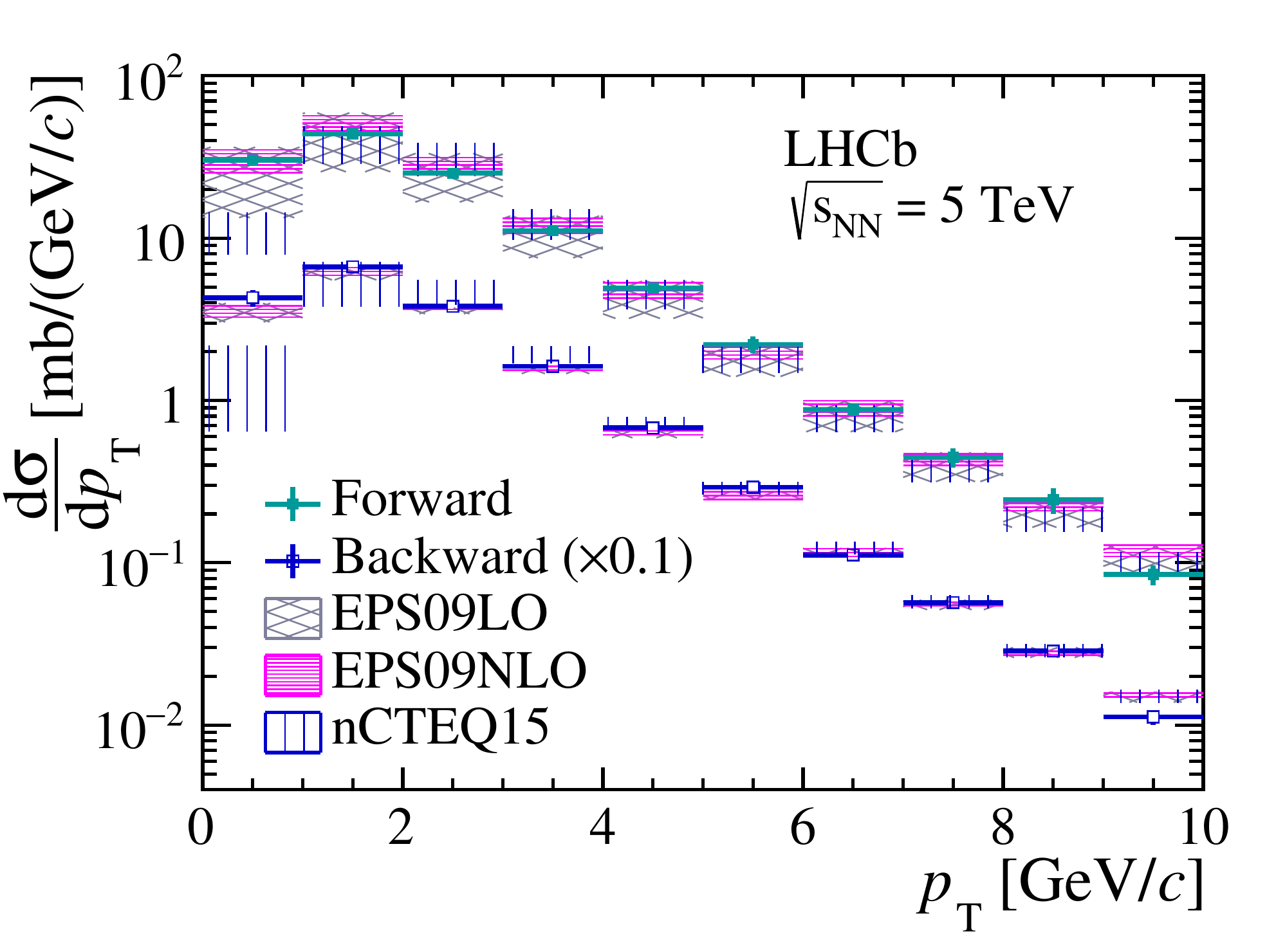}
\end{minipage}
\begin{minipage}[t]{0.49\textwidth}
\centering
\includegraphics[width=1.0\textwidth]{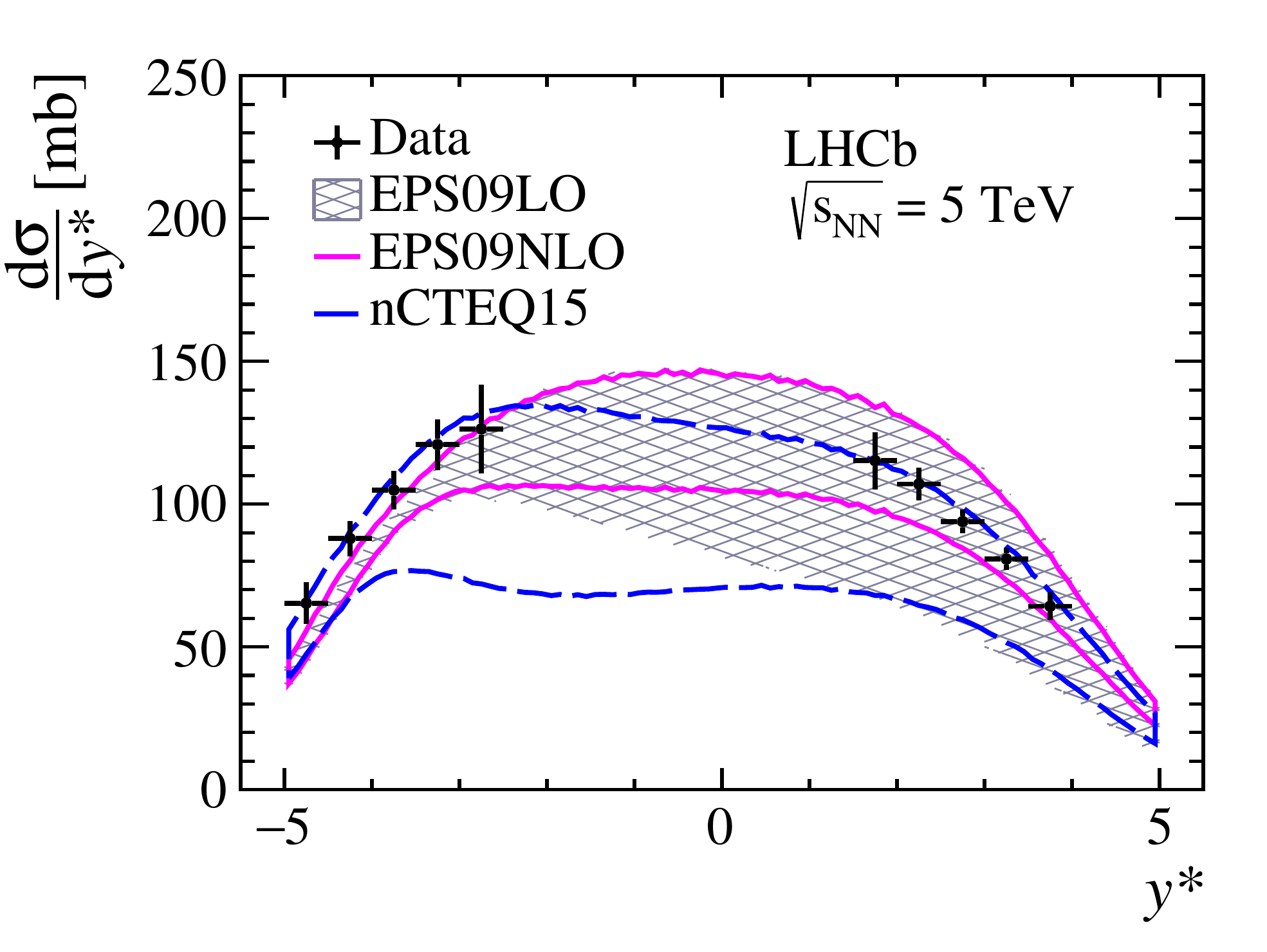}
\end{minipage}
\caption{Differential cross-section of prompt $\Dz$ meson production in $\pPb$ collisions as a function of (left) $\pt$ ($\frac{\deriv\sigma}{\deriv p_{\rm T}}$) and
(right) $y^{*}$ ($\frac{\deriv\sigma}{\deriv y^*}$) 
 in the forward and backward collision samples. 
The uncertainty is the quadratic sum of the statistical and systematic 
components.
The measurements are compared with theoretical predictions including 
different nuclear parton distribution functions as explained in the text.
}
\label{fig:CrossSection1D}
\end{figure}

The measured values of the double-differential cross-section of prompt 
$\Dz$ mesons in proton-lead collisions in the forward and backward regions
as a function of $\pt$ and $y^{*}$ are given in Table~\ref{tab:CrossSection2D} and shown in Fig.~\ref{fig:CrossSection2D}.
\begin{table}[!h]
\centering
\rotatebox{90}{%
\begin{varwidth}{0.95\textheight}
\caption{Double-differential cross-section $\frac{\deriv^2\sigma}{\deriv p_{\mathrm T}\deriv y^*}$\,(mb/(\gevc)) for prompt $\Dz$ meson
production as functions 
of $\pt$ and $y^{*}$ in $\pPb$ forward
and backward data, respectively. The first uncertainty is
statistical, the second is the component of the systematic uncertainty that 
is uncorrelated between bins and the
third is the correlated component. In the regions with no entries the signal
is not statistically significant.}
\label{tab:CrossSection2D}
\scalebox{0.78}{
\begin{tabular}{cccccc}
\hline
&\multicolumn{5}{c}{Forward (mb/(\gevc))}\\
$\pt[\gevc]$ &$1.5<y^{*}< 2.0$&$2.0<y^{*}<2.5$ &$2.5<y^{*}< 3.0$&$3.0<y^{*}<3.5$&$3.5<y^{*}< 4.0$\\
\hline
$[0,1]$ &$24.67\pm 0.32\pm 0.50\pm 3.45$&$23.48\pm 0.17\pm 0.25\pm 1.70$ &$22.01\pm 0.16\pm 0.20\pm 1.16$	&$20.19\pm 0.21\pm 0.23\pm 1.02$	&$18.41\pm 0.36\pm 0.33\pm 1.09$	\\
$[1,2]$ &$40.79\pm 0.34\pm 0.61\pm 3.83$&$38.45\pm 0.19\pm 0.35\pm 2.19$ &$33.79\pm 0.18\pm 0.26\pm 1.50$	&$29.89\pm 0.22\pm 0.28\pm 1.31$	&$24.17\pm 0.34\pm 0.40\pm 1.63$	\\
$[2,3]$ &$25.50\pm 0.20\pm 0.39\pm 1.76$&$23.73\pm 0.11\pm 0.20\pm 1.08$ &$20.34\pm 0.10\pm 0.16\pm 0.82$	&$16.84\pm 0.11\pm 0.17\pm 0.69$	&$13.03\pm 0.17\pm 0.23\pm 0.78$	\\
$[3,4]$ &$12.46\pm 0.11\pm 0.21\pm 0.63$&$11.09\pm 0.06\pm 0.10\pm 0.47$ &$ \phantom{0}9.31\pm 0.05\pm 0.09\pm 0.38$	&$ \phantom{0}7.73\pm 0.06\pm 0.09\pm 0.36$	&$\phantom{0}5.22\pm 0.09\pm 0.11\pm 0.46$	\\
$[4,5]$ &$\phantom{0} 5.79\pm 0.06\pm 0.11\pm 0.27$&$\phantom{0} 5.23\pm 0.04\pm 0.06\pm 0.21$ &$\phantom{0} 4.36\pm 0.03\pm 0.05\pm 0.17$	&$\phantom{0} 3.32\pm 0.04\pm 0.05\pm 0.14$	&$\phantom{0} 2.17\pm 0.07\pm 0.07\pm 0.45$	\\
$[5,6]$ &$\phantom{0} 2.94\pm 0.04\pm 0.07\pm 0.14$&$\phantom{0} 2.53\pm 0.03\pm 0.04\pm 0.11$ &$\phantom{0} 2.04\pm 0.02\pm 0.03\pm 0.09$	&$\phantom{0} 1.47\pm 0.02\pm 0.03\pm 0.10$	&$\phantom{0} 0.93\pm 0.07\pm 0.07\pm 0.37$	\\
$[6,7]$ &$\phantom{0} 1.42\pm 0.02\pm 0.04\pm 0.08$&$\phantom{0} 1.26\pm 0.02\pm 0.02\pm 0.05$ &$\phantom{0} 1.04\pm 0.02\pm 0.02\pm 0.06$	&$\phantom{0} 0.72\pm 0.02\pm 0.02\pm 0.10$	&$\phantom{0} 0.31\pm 0.08\pm 0.06\pm 0.20$	\\
$[7,8]$ &$\phantom{0} 0.84\pm 0.02\pm 0.03\pm 0.04$&$\phantom{0} 0.66\pm 0.01\pm 0.02\pm 0.04$ &$\phantom{0} 0.53\pm 0.01\pm 0.01\pm 0.03$	&$\phantom{0} 0.36\pm 0.02\pm 0.02\pm 0.09$	&$-$	\\
$[8,9]$ &$\phantom{0} 0.47\pm 0.01\pm 0.02\pm 0.02$&$\phantom{0} 0.38\pm 0.01\pm 0.01\pm 0.03$ &$\phantom{0} 0.32\pm 0.01\pm 0.01\pm 0.03$	&$\phantom{0}0.17\pm 0.02\pm 0.02\pm 0.06$	&$ -$	\\
$[9,10]$&$\phantom{0} 0.31\pm 0.01\pm 0.02\pm 0.02$&$\phantom{0} 0.24\pm 0.01\pm 0.01\pm 0.02$ &$\phantom{0} 0.17\pm 0.01\pm 0.01\pm 0.02$	&$\phantom{0}0.07\pm 0.01\pm 0.01\pm 0.03$	&$ -$	\\
\hline
&\multicolumn{5}{c}{Backward (mb/(\gevc))}\\
$\pt[\gevc]$ &$-3.0<y^{*}<-2.5$&$-3.5<y^{*}<-3.0$ &$-4.0<y^{*}<-3.5$&$-4.5<y^{*}<-4.0$&$-5.0<y^{*}<-4.5$\\
\hline
$[0,1]$ &$27.75\pm 0.48\pm 0.47\pm 5.78$	&$29.56\pm 0.33\pm 0.29\pm 2.98$	&$28.47\pm 0.38\pm 0.28\pm 1.98$	&$25.03\pm 0.58\pm 0.28\pm 1.78$	&$20.85\pm 1.08\pm 0.43\pm 2.21$	\\
$[1,2]$ &$46.66\pm 0.51\pm 0.69\pm 6.13$	&$46.10\pm 0.35\pm 0.38\pm 3.40$	&$40.35\pm 0.38\pm 0.33\pm 2.61$	&$35.82\pm 0.56\pm 0.38\pm 2.54$	&$27.00\pm 1.01\pm 0.45\pm 2.81$	\\
$[2,3]$ &$28.55\pm 0.29\pm 0.41\pm 2.41$	&$25.90\pm 0.19\pm 0.22\pm 1.62$	&$21.47\pm 0.18\pm 0.17\pm 1.26$	&$17.13\pm 0.23\pm 0.19\pm 1.09$	&$11.82\pm 0.45\pm 0.23\pm 0.97$	\\
$[3,4]$ &$12.73\pm 0.15\pm 0.18\pm 0.93$	&$10.98\pm 0.10\pm 0.10\pm 0.64$	&$\phantom{0} 8.75\pm 0.09\pm 0.08\pm 0.50$	&$\phantom{0} 6.33\pm 0.10\pm 0.08\pm 0.45$	&$ \phantom{0}3.61\pm 0.17\pm 0.09\pm 0.55$	\\
$[4,5]$ &$\phantom{0} 5.60\pm 0.08\pm 0.09\pm 0.38$&$\phantom{0} 4.59\pm 0.05\pm 0.05\pm 0.26$&$\phantom{0} 3.36\pm 0.05\pm 0.04\pm 0.19$&$\phantom{0}2.21\pm 0.05\pm 0.03\pm 0.14$	&$ \phantom{0}1.47\pm 0.13\pm 0.06\pm 0.43$	\\
$[5,6]$ &$\phantom{0} 2.53\pm 0.05\pm 0.05\pm 0.16$&$\phantom{0} 1.93\pm 0.03\pm 0.03\pm 0.11$&$\phantom{0} 1.38\pm 0.03\pm 0.02\pm 0.08$&$\phantom{0}0.82\pm 0.03\pm 0.02\pm 0.10$	&$ \phantom{0}0.57\pm 0.14\pm 0.06\pm 0.30$	\\
$[6,7]$ &$\phantom{0} 1.32\pm 0.03\pm 0.03\pm 0.08$&$\phantom{0} 0.92\pm 0.02\pm 0.02\pm 0.06$&$\phantom{0} 0.62\pm 0.02\pm 0.01\pm 0.04$&$\phantom{0}0.28\pm 0.02\pm 0.01\pm 0.07$	&$ -$	\\
$[7,8]$ &$\phantom{0} 0.65\pm 0.02\pm 0.02\pm 0.04$&$\phantom{0} 0.48\pm 0.02\pm 0.01\pm 0.04$&$\phantom{0} 0.31\pm 0.01\pm 0.01\pm 0.04$&$\phantom{0}0.19\pm 0.03\pm 0.01\pm 0.08$	&$ -$	\\
$[8,9]$ &$\phantom{0} 0.33\pm 0.02\pm 0.01\pm 0.02$&$\phantom{0} 0.24\pm 0.01\pm 0.01\pm 0.02$&$\phantom{0} 0.14\pm 0.01\pm 0.01\pm 0.03$&$\phantom{0}0.11\pm 0.03\pm 0.01\pm 0.08$	&$ -$	\\
$[9,10]$&$\phantom{0} 0.22\pm 0.01\pm 0.01\pm 0.02$&$\phantom{0} 0.13\pm 0.01\pm 0.01\pm 0.01$&$\phantom{0} 0.08\pm 0.01\pm 0.00\pm 0.02$&$-$	&$ -$	\\

\hline
\end{tabular}
}\end{varwidth}
}
\end{table}
The one-dimensional differential prompt $\Dz$ meson cross-sections as a function of $\pt$ or $y^{*}$ are reported
 in Tables~\ref{tab:CrossSection1DPT} and \ref{tab:CrossSection1DY}, and are displayed in Fig.~\ref{fig:CrossSection1D}. 
The measurements are also shown as a function of $\pt$ integrated\footnote{
The integration over $y^{\ast}$ is 
performed up to $|y^{\ast}|$=3.5 for $\pt > 6\gevc$, neglecting the bin
 $3.5<|y^{*}|<4.0$ since it is not populated
in the forward sample.
This applies for the integrated cross-sections presented in
this subsection, in Tables~\ref{tab:CrossSection1DPT}, \ref{tab:RpPbResult} and \ref{tab:RFBResult} and in Figs.~\ref{fig:CrossSection1D}, \ref{fig:RpPbPT}, \ref{fig:RFBResult} and \ref{fig:RFBD0Jpsi}.}
over
$y^{*}$ in the common rapidity range $2.5<|y^{*}|<4.0$.
\newline
\begin{table}[btp]
\caption{Measured differential cross-section $\frac{\deriv\sigma}{\deriv p_{\mathrm T}}$\,(mb/(\gevc))  for prompt $\Dz$ meson production as a function of $\pt$ in $\pPb$ forward
and backward data, respectively. The first uncertainty is
statistical, the second is the component of the systematic uncertainty that is uncorrelated between bins and the
third is the correlated component. The results in the last two columns are 
integrated over the common rapidity range $2.5<|y^{*}|<4.0$ for $\pt<6\gevc$ and over $2.5<|y^{*}|<3.5$ for $6<\pt<10\gevc$.
}
\centering
\scalebox{0.9}{
\begin{tabular}{cccc}
\hline
&\multicolumn{3}{c}{Forward (mb/(\gevc))}\\
$\pt[\gevc]$ &$1.5<y^{*}< 4.0$&$2.5<y^{*}<4.0$ &$2.5<y^{*}<3.5$\\
\hline
$[0,1]$ &$54.38\pm 0.29\pm 0.36\pm 3.96$	        &$30.31\pm 0.22\pm 0.22\pm 1.59$             &$-$\\
$[1,2]$ &$83.54\pm 0.30\pm 0.45\pm 5.01$	        &$43.92\pm 0.22\pm 0.28\pm 2.17$             &$-$\\
$[2,3]$ &$49.72\pm 0.16\pm 0.27\pm 2.45$	        &$25.11\pm 0.11\pm 0.16\pm 1.11$             &$-$\\
$[3,4]$ &$22.91\pm 0.09\pm 0.14\pm 1.10$	        &$11.13\pm 0.06\pm 0.08\pm 0.55$             &$-$\\
$[4,5]$ &$10.43\pm 0.06\pm 0.08\pm 0.54$	        &$\phantom{0} 4.92\pm 0.04\pm 0.05\pm 0.32$  &$-$\\
$[5,6]$ &$\phantom{0} 4.95\pm 0.05\pm 0.06\pm 0.35$	&$\phantom{0} 2.21\pm 0.04\pm 0.04\pm 0.26$  &$-$\\
$[6,7]$ &$\phantom{0} 2.37\pm 0.05\pm 0.04\pm 0.21$	&$-$&  $\phantom{0} 0.88\pm 0.01\pm 0.01\pm 0.07$\\
$[7,8]$ &$\phantom{0} 1.20\pm 0.02\pm 0.02\pm 0.09$	&$-$&  $\phantom{0} 0.45\pm 0.01\pm 0.01\pm 0.06$\\
$[8,9]$ &$\phantom{0} 0.67\pm 0.01\pm 0.01\pm 0.06$	&$-$&  $\phantom{0} 0.24\pm 0.01\pm 0.01\pm 0.04$\\
$[9,10]$&$\phantom{0} 0.39\pm 0.01\pm 0.01\pm 0.04$	&$-$&  $\phantom{0} 0.08\pm 0.00\pm 0.00\pm 0.01$\\
\hline
&\multicolumn{3}{c}{Backward (mb/(\gevc))}\\
$\pt[\gevc]$ &$-5.0<y^{*}<-2.5$&$-4.0<y^{*}<-2.5$ &$-3.5<y^{*}<-2.5$\\
\hline
$[0,1]$ &$65.83\pm 0.70\pm 0.40\pm 6.85$	        &$42.89\pm 0.35\pm 0.31\pm 5.15$            &$-$\\
$[1,2]$ &$97.97\pm 0.68\pm 0.52\pm 8.30$	        &$66.56\pm 0.36\pm 0.43\pm 5.80$            &$-$\\
$[2,3]$ &$52.43\pm 0.32\pm 0.29\pm 3.57$	        &$37.96\pm 0.20\pm 0.25\pm 2.56$            &$-$\\
$[3,4]$ &$21.21\pm 0.14\pm 0.13\pm 1.45$	        &$16.23\pm 0.10\pm 0.11\pm 1.01$            &$-$\\
$[4,5]$ &$\phantom{0} 8.62\pm 0.09\pm 0.06\pm 0.62$	&$\phantom{0} 6.78\pm 0.05\pm 0.05\pm 0.41$ &$-$\\
$[5,6]$ &$\phantom{0} 3.61\pm 0.08\pm 0.04\pm 0.33$	&$\phantom{0} 2.92\pm 0.03\pm 0.03\pm 0.18$ &$-$\\
$[6,7]$ &$\phantom{0} 1.57\pm 0.03\pm 0.02\pm 0.12$	&$-$ &$\phantom{0} 1.12\pm 0.02\pm 0.02\pm 0.07$\\
$[7,8]$ &$\phantom{0} 0.81\pm 0.02\pm 0.01\pm 0.09$	&$-$ &$\phantom{0} 0.57\pm 0.01\pm 0.01\pm 0.04$\\
$[8,9]$ &$\phantom{0} 0.41\pm 0.02\pm 0.01\pm 0.07$	&$-$ &$\phantom{0} 0.29\pm 0.01\pm 0.01\pm 0.02$\\
$[9,10]$&$\phantom{0} 0.22\pm 0.01\pm 0.01\pm 0.02$	&$-$ &$\phantom{0} 0.11\pm 0.01\pm 0.01\pm 0.01$\\

\hline
\end{tabular}
}
\label{tab:CrossSection1DPT}
\end{table}
\begin{table}[btp]
\caption{Differential cross-section $\frac{\deriv\sigma}{\deriv y^*}$\,(mb) for prompt $\Dz$ meson production as a function of $|y^{*}|$ in $\pPb$ forward
and backward data, respectively. The first uncertainty is
statistical, the second is the component of the systematic uncertainty that is uncorrelated between bins and the
third is the correlated component.}
\centering
\begin{tabular}{cc}
\hline
\multicolumn{2}{c}{Forward (mb)}\\
$y^{*}$ &$0<\pt<10\gevc$\\
\hline
$[1.5,2.0]$&$115.19\pm 0.53\pm 0.91\pm 9.99$\\
$[2.0,2.5]$&$107.05\pm 0.29\pm 0.50\pm 5.73$\\
$[2.5,3.0]$&$\phantom{0}93.90\pm 0.27\pm 0.38\pm 4.14$\\
$[3.0,3.5]$&$\phantom{0}80.76\pm 0.33\pm 0.42\pm 3.71$\\
$[3.5,4.0]$&$\phantom{0}64.24\pm 0.55\pm 0.58\pm 4.79$\\
\hline
\multicolumn{2}{c}{Backward (mb)}\\
$y^{*}$ &$0<\pt<10\gevc$\\
\hline
$[-3.0,-2.5]$&$126.35\pm 0.78\pm 0.95\pm15.54$\\
$[-3.5,-3.0]$&$120.84\pm 0.53\pm 0.53\pm \phantom{0}8.89$\\
$[-4.0,-3.5]$&$104.93\pm 0.58\pm 0.47\pm \phantom{0}6.66$\\
$[-4.5,-4.0]$&$\phantom{0}87.92\pm 0.85\pm 0.52\pm \phantom{0}6.13$\\
$[-5.0,-4.5]$&$\phantom{0}65.32\pm 1.57\pm 0.68\pm \phantom{0}7.07$\\
\hline
\end{tabular}
\label{tab:CrossSection1DY}
\end{table}
The integrated cross-sections of prompt $\Dz$ meson production in $\pPb$ forward data in the full and common fiducial regions
are
\begin{equation}
\sigma_\mathrm{forward}(\pt<10\gevc, 1.5<y^{*}<4.0) =230.6\pm0.5\pm13.0\mbarn, \nonumber
\end{equation}
\begin{equation}
\sigma_\mathrm{forward}(\pt<10\gevc, 2.5<y^{*}<4.0) =119.1\pm0.3\pm\phantom{0}5.6\mbarn. \nonumber
\end{equation}
The integrated cross-sections of prompt $\Dz$ meson production in Pb$p$ backward data in the two fiducial regions
are
\begin{equation}
\sigma_\mathrm{backward}(\pt<10\gevc, -2.5<y^{*}<-5.0) =252.7\pm1.0\pm20.0\mbarn, \nonumber
\end{equation}
\begin{equation}
\sigma_\mathrm{backward}(\pt<10\gevc, -2.5<y^{*}<-4.0) =175.5\pm0.6\pm14.4\mbarn, \nonumber
\end{equation}
where the first uncertainties are statistical and the second systematic.

The cross-sections as a function of $\pt$ and $y^*$, shown in 
Fig.~\ref{fig:CrossSection1D}, are compared with calculations~\cite{ShaoN1,ShaoN2,ShaoN3}
validated with results 
 of heavy-flavour production cross-section in $pp$ collisions.
The nuclear effects are considered by using three different sets of nuclear parton distribution functions (nPDFs), 
 the leading-order EPS09 (EPS09LO)~\cite{EPS09}, the next-to-leading order EPS09 (EPS09NLO)~\cite{EPS09} and nCTEQ15~\cite{nCTEQ15}.
The free nucleon PDF CT10NLO~\cite{protonPDF} is also used as a reference for the cross-section predictions in $pp$ collisions.
Within large theoretical uncertainties, all three sets of nPDFs can 
give descriptions consistent with the \lhcb data, although a discrepancy is observed in the low $\pt$ region 
between the measurements and the nCTEQ15 predictions.
\subsection{Nuclear modification factors}

\begin{figure}[tbp]
\centering
\begin{minipage}[t]{0.49\textwidth}
\centering
\includegraphics[width=1.0\textwidth]{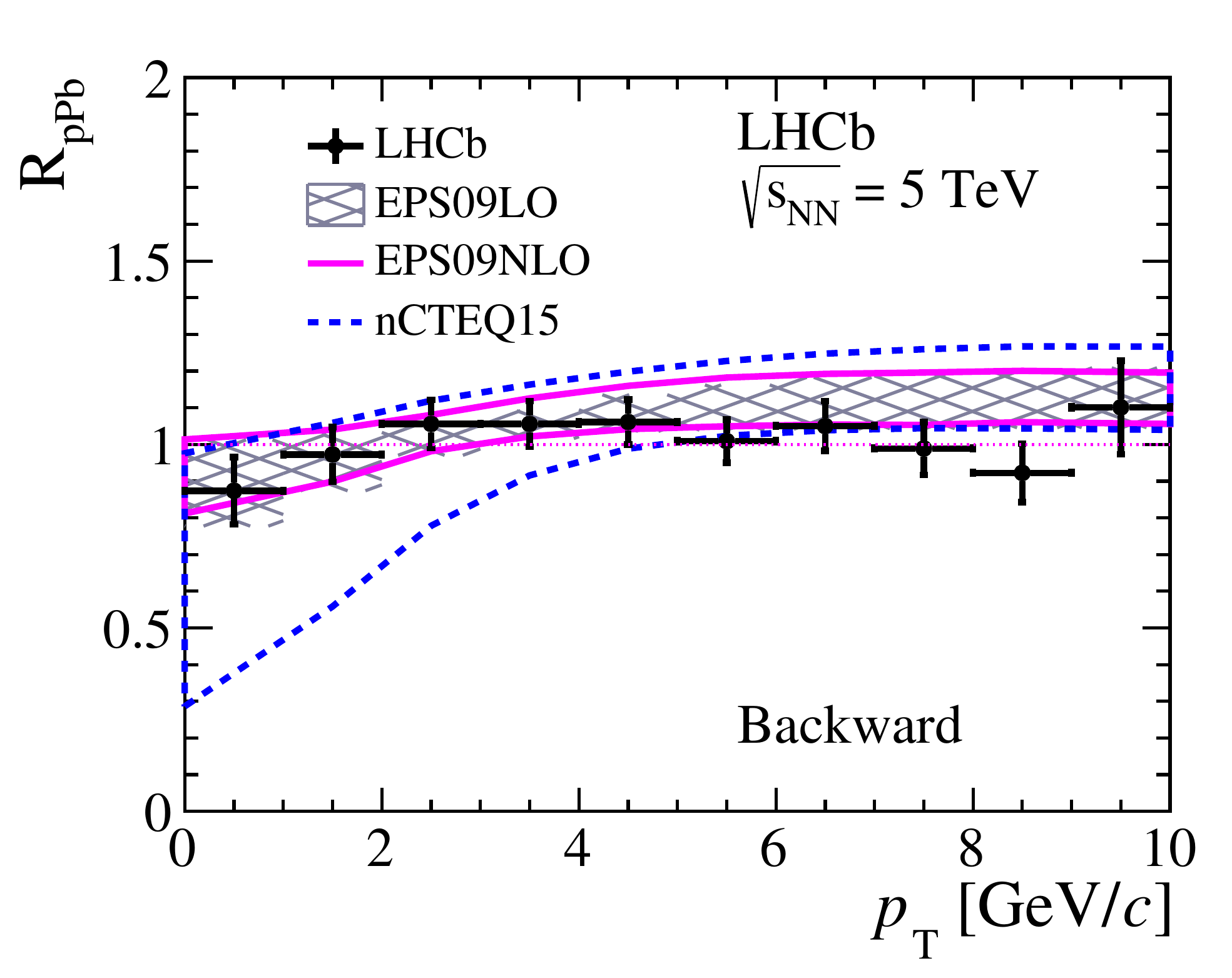}
\end{minipage}
\begin{minipage}[t]{0.49\textwidth}
\centering
\includegraphics[width=1.0\textwidth]{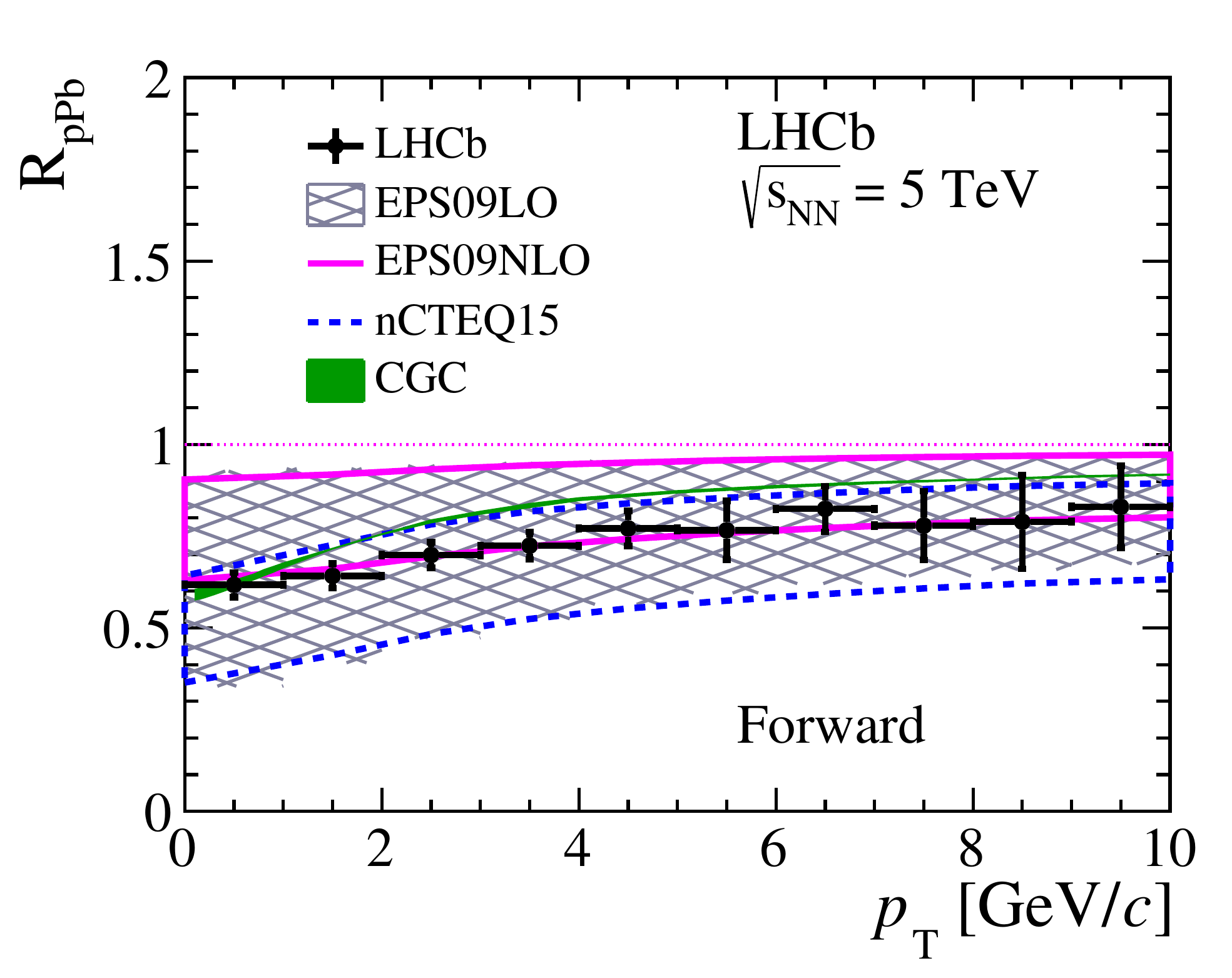}
\end{minipage}
\caption{Nuclear modification factor $R_{\pPb}$ as a function of $\pt$ for prompt $\Dz$ meson
production in the (left) backward data and (right) forward data, 
integrated over the common rapidity range $2.5<|y^{*}|<4.0$ for $\pt<6\gevc$ and over $2.5<|y^{*}|<3.5$ for $6<\pt<10\gevc$.
The uncertainty is the quadratic sum of the statistical and systematic components. The CGC
 predictions are only available for the forward region.
}
\label{fig:RpPbPT}
\end{figure}

\begin{figure}[tbp]
\centering
\begin{minipage}[t]{0.7\textwidth}
\centering
\includegraphics[width=\textwidth]{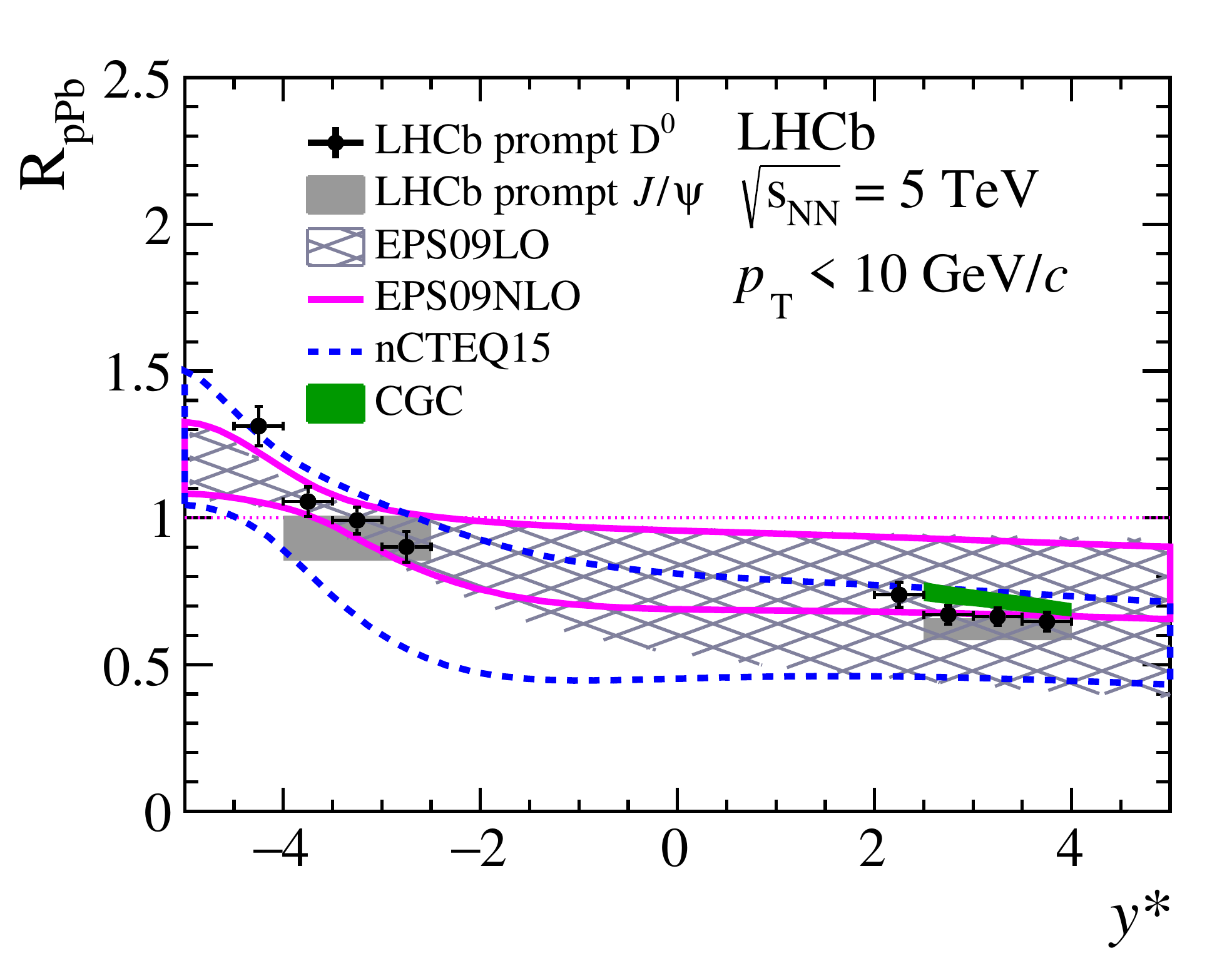}
\end{minipage}
\caption{Nuclear modification factor $R_{\pPb}$ as a function of $y^*$ for prompt $\Dz$ meson production, integrated  up to  $\pt=10\gevc$. 
The uncertainty is the quadratic sum of the statistical and systematic components.
}
\label{fig:RpPbY}
\end{figure}

The value of the $\Dz$ meson production cross-section in $pp$
collisions at $5\tev$, needed
for the measurement of the nuclear modification factor $R_{\pPb}$,
is taken from the LHCb measurement~\cite{LHCb-PAPER-2016-042}.
Correlations between the uncertainties of quantities that are common to both measurements are taken into account.
The nuclear modification
factor for prompt $\Dz$ meson production is shown in Fig.~\ref{fig:RpPbPT} in bins of $\pt$ and Fig.~\ref{fig:RpPbY} in bins of $y^*$.  
The nuclear modification factors are calculated as a function of $\pt$ integrated over $y^*$ in the ranges described in Fig.~\ref{fig:RpPbPT} for both
forward and backward samples.
The values of $R_{\pPb}$, summarised in Tables~\ref{tab:RpPbResult} and~\ref{tab:RpPbResultY},
show a slight increase as a function of  $\pt$, suggesting 
that the suppression may decrease with increasing transverse momentum.
%

The measurements are compared with calculations using EPS09LO, EPSNLO and nCTEQ15 nPDFs~\cite{ShaoN2,ShaoN1,ShaoN3}. For
the results in the backward configuration, all three predictions show reasonable agreement with each other and with LHCb
data. In the forward configuration, nCTEQ15 and EPS09LO show better agreement with the data than EPS09NLO.
Calculations~\cite{MNR} using  CTEQ6M~\cite{CTEQ6M} nucleon PDF and EPS09NLO nPDF give results for $R_{\pPb}$ that are similar 
to a combination of CT10NLO and EPS09NLO.

\begin{figure}[tbp]
\centering
\includegraphics[width=0.7\textwidth]{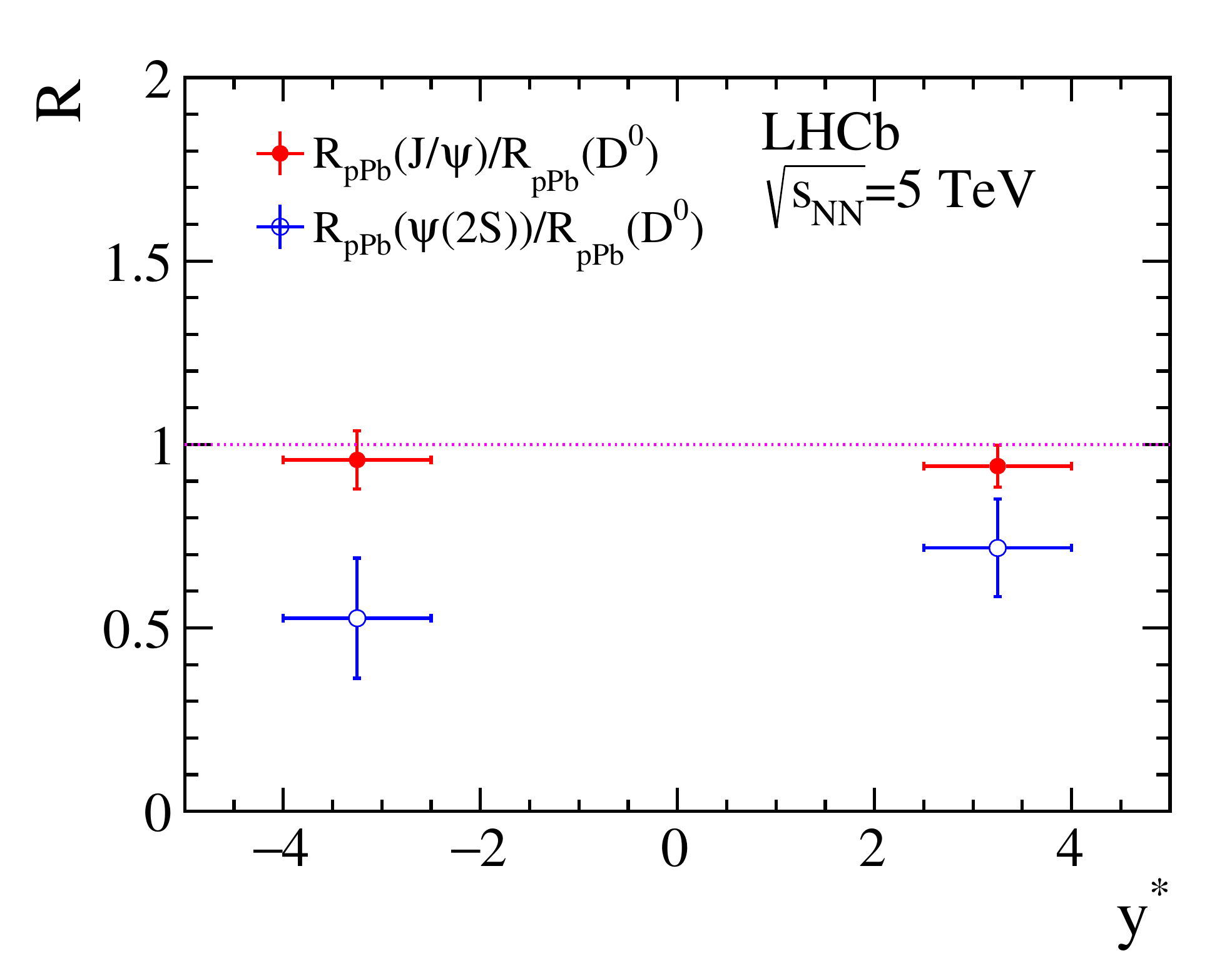}
\caption{Ratio of nuclear modification factors $R_{\pPb}$ of
 $\jpsi$ and $\psitwos$ to $\Dz$ mesons 
in bins of rapidity integrated up to $\pt=10\gevc$ in the common rapidity range 
 $2.5<|y^{*}|<4.0$.
The uncertainty is the quadratic sum of the statistical and systematic components.
}
\label{fig:RpPbDoubleRatio}
\end{figure}

The nuclear modification factors for prompt $D^0$ are also compared with those for prompt
$\jpsi$~\cite{LHCb-PAPER-2013-052} in Fig.~\ref{fig:RpPbY} as a function of
$\pt$ integrated over rapidity, and they are found to be consistent.
This is the first measurement of $R_{\pPb}$ in this kinematic range. 
The ratios of the nuclear modification factors of  $\jpsi$ and 
$\psitwos$ mesons to $\Dz$ mesons as a function of rapidity are shown in Fig.~\ref{fig:RpPbDoubleRatio}
where a different suppression between the two charmonium states can be 
observed.
In 
Figs.~\ref{fig:RpPbPT} and ~\ref{fig:RpPbY}  the measurements are also compared with calculations 
in the colour glass condensate
framework (CGC)~\cite{Ducloue:2015gfa}, which includes the effect of the saturation of partons at 
small $x$.
The CGC model is found to be able to describe the trend of prompt $\Dz$ meson nuclear modifications as a function
 of $\pt$ and of rapidity. The uncertainty band for this model is much smaller than for the 
nuclear PDF calculations, since it only contains the variation of charm quark masses and 
factorisation scale which largely cancel in this ratio of cross-sections.
Another CGC framework calculation gives similar results for nuclear modifications of charm
production~\cite{Fujii:2017rqa}.
In the context of $p$Pb collisions, recent measurements have shown 
that long-range collective effects, which have previously been 
observed in relatively large nucleus-nucleus collision systems, 
may also be present in smaller collision systems at large charged 
particle multiplicities~\cite{CMS:2012qk,Abelev:2012ola,Aad:2012gla,Adare:2013piz}. 
If these effects are due to the creation of a hydrodynamic system, 
momentum anisotropies at 
the quark level can arise, which may modify the final distribution 
of observed heavy-quark hadrons~\cite{Beraudo:2015wsd}. Since the 
measurements in this analysis do not consider a classification in 
charged particle multiplicity, potential modifications in high-multiplicity
 events are weakened as the presented observables 
are integrated over charged particle multiplicity.
\begin{table}[btp]
\caption{Nuclear modification factor $R_{\pPb}$ for prompt $\Dz$ meson production
 in different $\pt$ ranges, 
integrated over the common rapidity range $2.5<|y^{*}|<4.0$ for $\pt<6\gevc$ and over $2.5<|y^{*}|<3.5$ for $6<\pt<10\gevc$
for the forward (positive $y^*$) and backward (negative $y^*$) samples.
The first uncertainty is statistical and the second systematic. 
}
\centering
\begin{tabular}{ccc}
\hline
$\pt[\gevc]$ &Forward & Backward\\
\hline
$[0,1]$&$ 0.62\pm 0.01\pm 0.03$	&$ 0.87\pm 0.01\pm 0.09$\\
$[1,2]$&$ 0.64\pm 0.01\pm 0.03$	&$ 0.97\pm 0.01\pm 0.07$\\
$[2,3]$&$ 0.70\pm 0.01\pm 0.03$	&$ 1.06\pm 0.01\pm 0.07$\\
$[3,4]$&$ 0.72\pm 0.01\pm 0.04$	&$ 1.06\pm 0.01\pm 0.06$\\
$[4,5]$&$ 0.77\pm 0.01\pm 0.05$	&$ 1.06\pm 0.01\pm 0.06$\\
$[5,6]$&$ 0.77\pm 0.02\pm 0.08$	&$ 1.01\pm 0.02\pm 0.06$\\
$[6,7]$&$ 0.82\pm 0.02\pm 0.06$	&$ 1.05\pm 0.03\pm 0.06$\\
$[7,8]$&$ 0.78\pm 0.03\pm 0.09$	&$ 0.99\pm 0.04\pm 0.06$\\
$[8,9]$&$ 0.79\pm 0.05\pm 0.12$	&$ 0.92\pm 0.05\pm 0.07$\\
$[9,10]$&$ 0.83\pm 0.07\pm 0.09$	&$ 1.10\pm 0.10\pm 0.09$\\
$[0,10]$&$ 0.66\pm 0.00\pm 0.03$    &$ 0.97\pm 0.01\pm 0.07$\\

\hline

\end{tabular}
\label{tab:RpPbResult}
\end{table}
\begin{table}[btp]
\caption{Nuclear modification factor $R_{\pPb}$ for prompt $\Dz$ meson production in different $y^*$ ranges, integrated  up to  $\pt=10\gevc$.
The first uncertainty is statistical and the second systematic. 
}
\centering
\begin{tabular}{ccc}
\hline
$y^{*}$ &$R_{p\mathrm{Pb}}$\\
\hline
$[-4.5,-4.0]$&$ 1.31\pm 0.02\pm 0.06$\\
$[-4.0,-3.5]$&$ 1.05\pm 0.01\pm 0.05$\\
$[-3.5,-3.0]$&$ 0.99\pm 0.01\pm 0.04$\\
$[-3.0,-2.5]$&$ 0.90\pm 0.01\pm 0.05$\\
$[2.0,2.5]$&$ 0.74\pm 0.01\pm 0.04$\\
$[2.5,3.0]$&$ 0.67\pm 0.00\pm 0.03$\\
$[3.0,3.5]$&$ 0.66\pm 0.00\pm 0.03$\\
$[3.5,4.0]$&$ 0.65\pm 0.01\pm 0.03$\\

\hline

\end{tabular}
\label{tab:RpPbResultY}
\end{table}
\subsection{Forward-backward ratio}

\begin{figure}[tbp]
\centering
\begin{minipage}[t]{0.49\textwidth}
\centering
\includegraphics[width=1.0\textwidth]{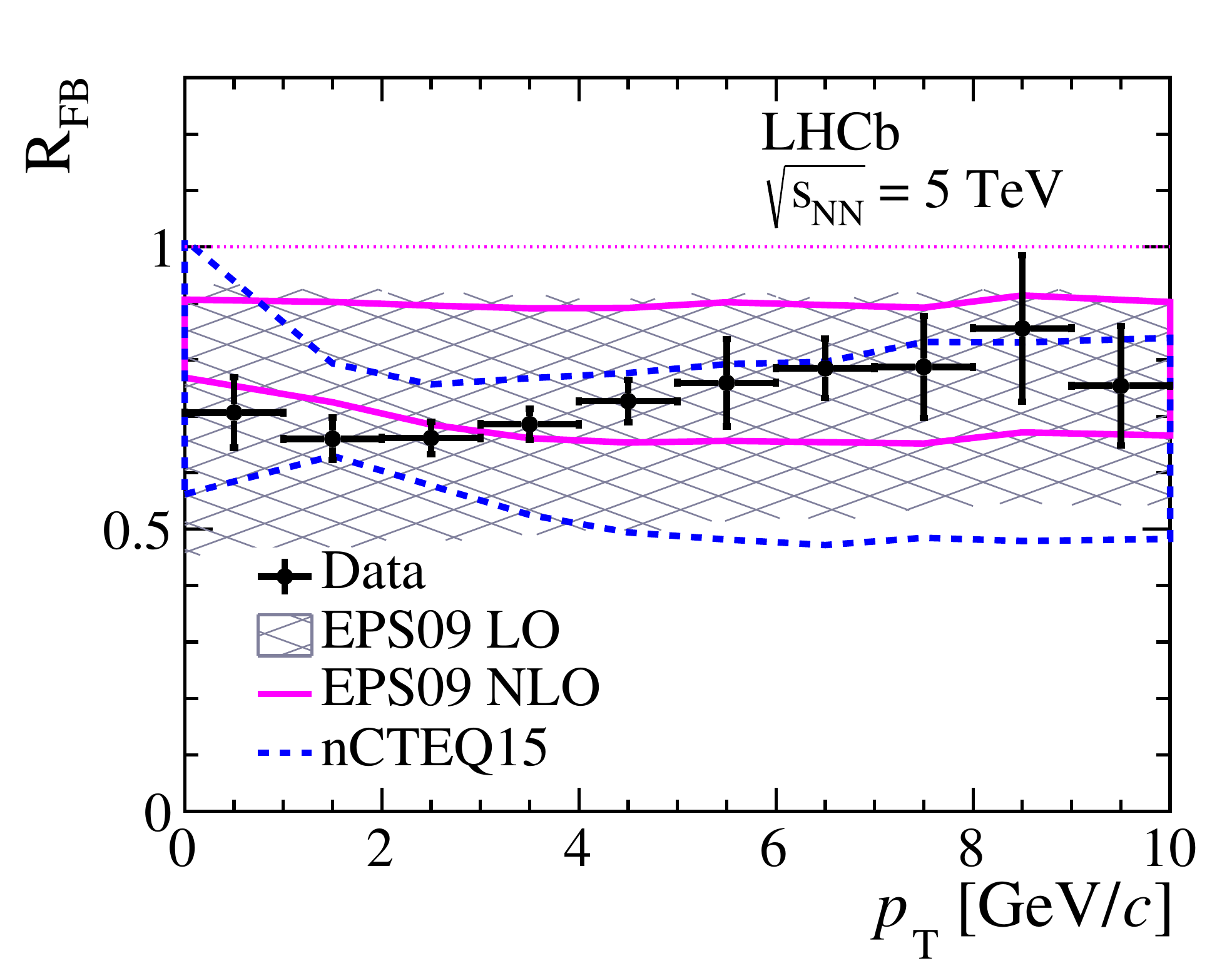}
\end{minipage}
\begin{minipage}[t]{0.49\textwidth}
\centering
\includegraphics[width=1.0\textwidth]{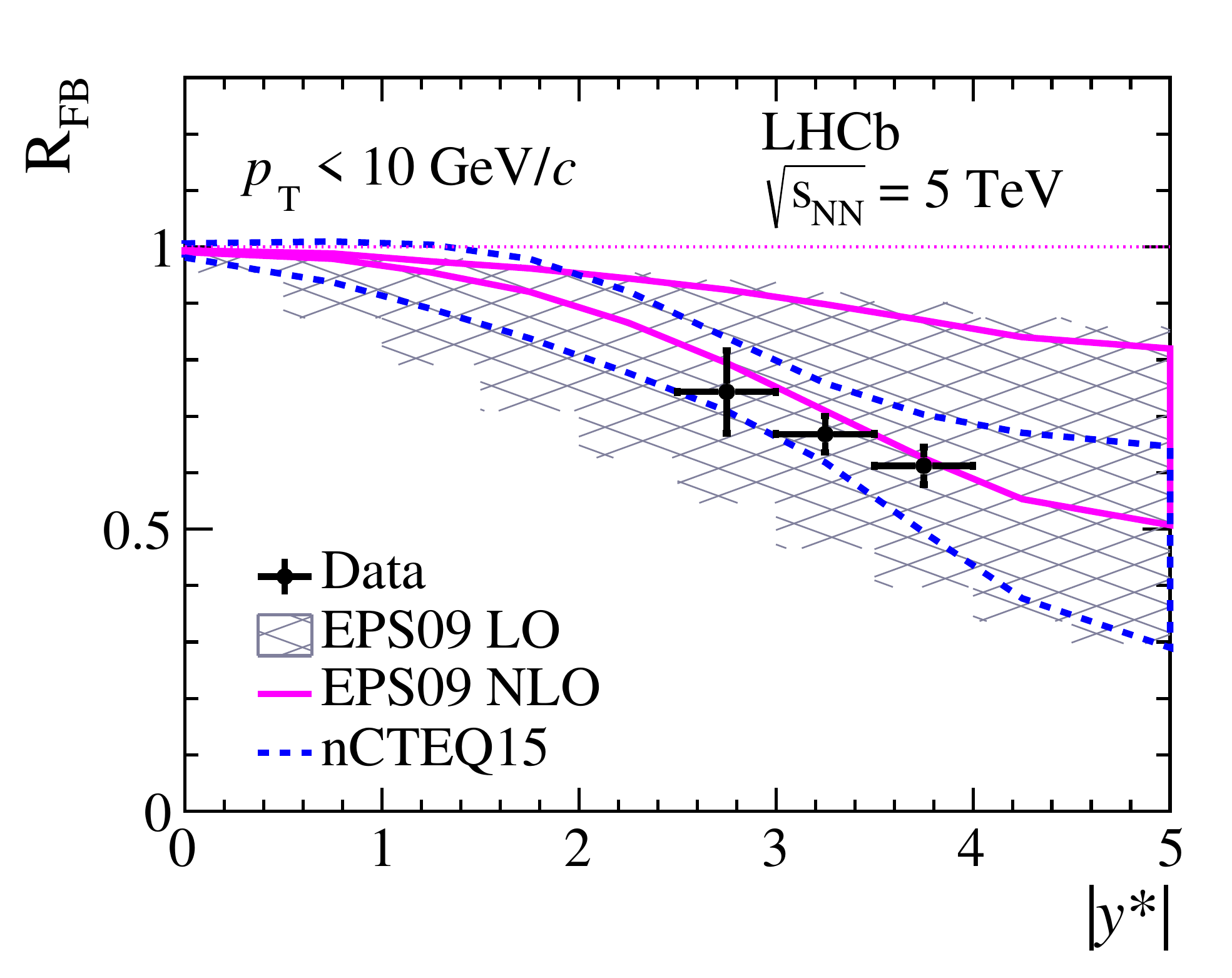}
\end{minipage}
\caption{Forward-backward ratio $R_\mathrm{FB}$ for prompt $\Dz$ meson 
production (left) as a function of $\pt$ integrated over the common rapidity 
range $2.5<|y^{*}|<4.0$ for $\pt<6\gevc$ and over $2.5<|y^{*}|<3.5$ for 
$6<\pt<10\gevc$; (right) as a function of $y^*$ integrated  up to  $\pt=10\gevc$.
The uncertainty is the quadratic sum of the statistical and systematic components.
} 
\label{fig:RFBResult}
\end{figure}

In the forward-backward production ratio $R_\mathrm{FB}$
the common uncertainty between the forward and backward measurements 
largely cancels.
The uncertainties of branching fraction, signal yield and tracking are considered fully correlated, while the PID
uncertainty is considered 90\% correlated since it is a mixture of statistical uncertainty (uncorrelated) and the uncertainties
due to the binning
scheme and yield determination (correlated). All other uncertainties are uncorrelated.  
The measured  $R_\mathrm{FB}$ values are shown in Fig.~\ref{fig:RFBResult}, as a function of 
$\pt$ integrated over the range $2.5<|y^*|<4.0$,
and as a function of $y^*$ integrated  up to  $\pt=10\gevc$.
 The $R_\mathrm{FB}$ values in different kinematic bins are also
summarised in Table~\ref{tab:RFBResult}. Good agreement is found between measurements and theoretical predictions using EPS09LO and nCTEQ15 nPDFs.

\begin{table}[btp]
\caption{Forward-backward ratio $R_\mathrm{FB}$ for prompt $\Dz$ meson 
production in different $\pt$ ranges, integrated over the common rapidity
 range $2.5<|y^{*}|<4.0$ for $\pt<6\gevc$ and over $2.5<|y^{*}|<3.5$ 
for $6<\pt<10\gevc$, and in different $y^*$ ranges  integrated  up to  
$\pt=10\gevc$.
The first uncertainty is the statistical and the second is the systematic component.
}
\centering
\begin{tabular}{cc}
\hline
$\pt[\gevc]$ &$R_\mathrm{FB}$\\
\hline
$[0,1]$&$ 0.71\pm 0.01\pm 0.06$\\
$[1,2]$&$ 0.66\pm 0.00\pm 0.04$\\
$[2,3]$&$ 0.66\pm 0.00\pm 0.03$\\
$[3,4]$&$ 0.69\pm 0.01\pm 0.03$\\
$[4,5]$&$ 0.73\pm 0.01\pm 0.04$\\
$[5,6]$&$ 0.76\pm 0.02\pm 0.08$\\
$[6,7]$&$ 0.79\pm 0.02\pm 0.05$\\
$[7,8]$&$ 0.79\pm 0.03\pm 0.09$\\
$[8,9]$&$ 0.86\pm 0.04\pm 0.12$\\
$[9,10]$&$ 0.75\pm 0.06\pm 0.09$\\
$[0,10]$&$ 0.68\pm 0.00\pm 0.04$\\
\hline
$|y^*|$ &$R_\mathrm{FB}$\\
\hline
$[2.5,3.0]$&$ 0.74\pm 0.01\pm 0.07$\\
$[3.0,3.5]$&$ 0.67\pm 0.00\pm 0.03$\\
$[3.5,4.0]$&$ 0.61\pm 0.01\pm 0.03$\\

\hline
\end{tabular}
\label{tab:RFBResult}
\end{table}
In the common kinematic range $\pt<10\gevc$, $2.5<|y^*|<4.0$, the forward-backward ratio 
$R_\mathrm{FB}$ is $0.71\pm 0.01(\mathrm{stat})\pm 0.04(\mathrm{syst})$, indicating  a significant
 asymmetry. 
The predictions for $R_\mathrm{FB}$ integrated over the same kinematic range are
 $0.71^{+0.21}_{-0.24}$ for EPS09 at leading order,
$0.81^{+0.10}_{-0.09}$ for EPS09 at next-to-leading order and $0.69^{+0.07}_{-0.07}$ for 
the nCTEQ15 nPDF set,
 which are all in good agreement with the measured value. 
The forward-backward production ratio increases slightly with increasing $\pt$, 
and decreases strongly 
with increasing rapidity $|y^*|$. This behaviour is consistent with the expectations 
from the QCD calculations.

\begin{figure}[tbp]
\centering
\begin{minipage}[t]{0.49\textwidth}
\centering
\includegraphics[width=1.0\textwidth]{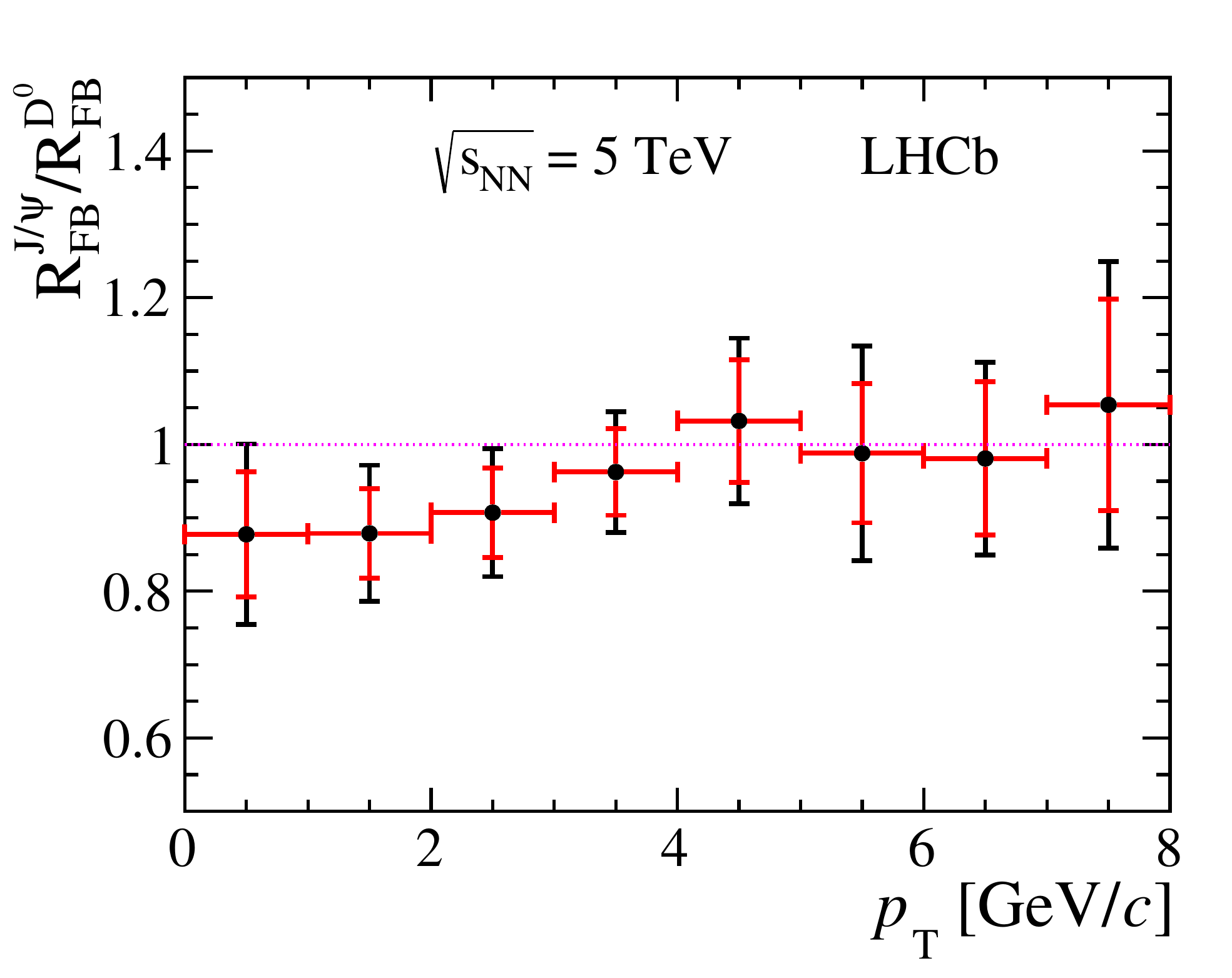}
\end{minipage}
\begin{minipage}[t]{0.49\textwidth}
\centering
\includegraphics[width=1.0\textwidth]{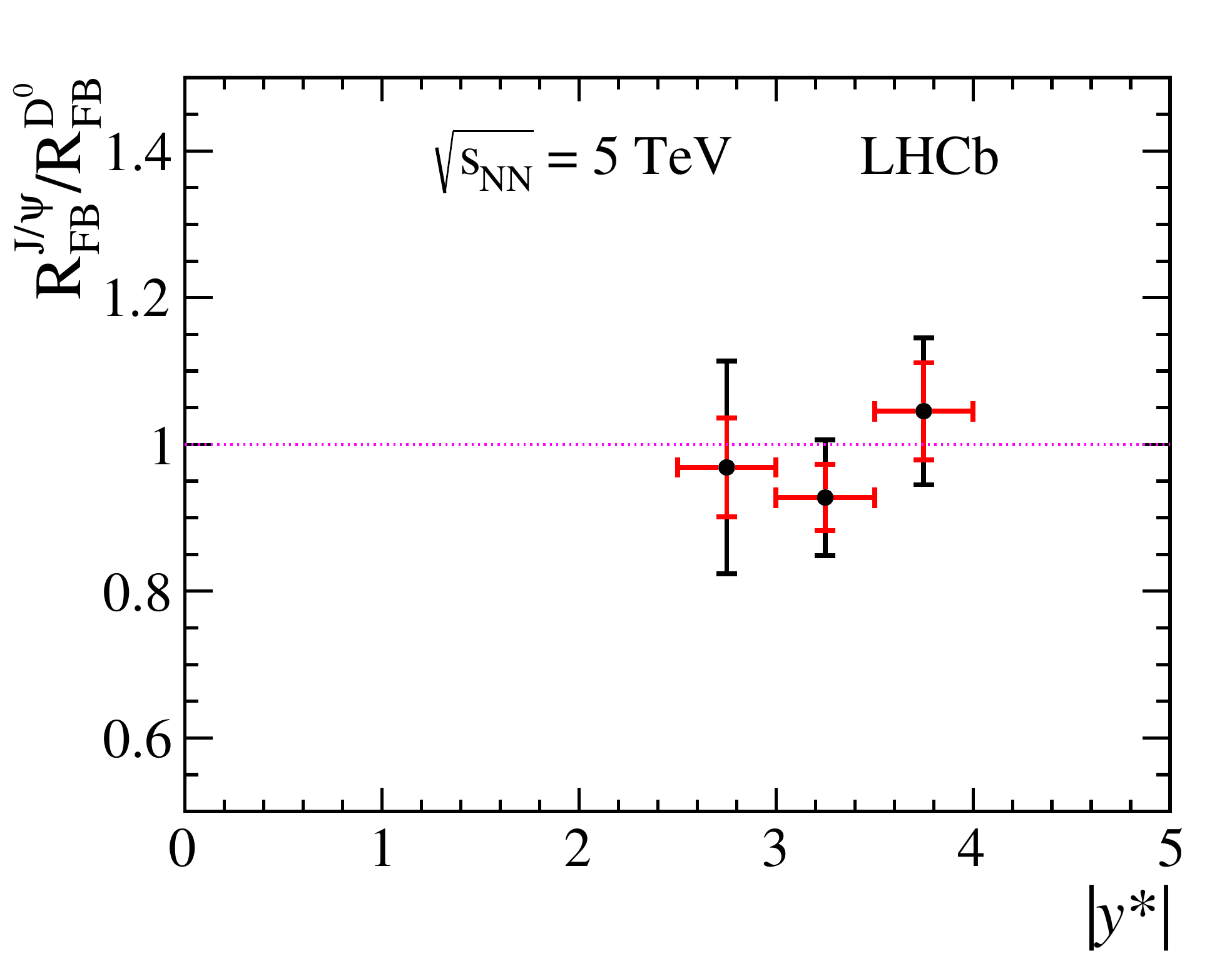}
\end{minipage}
\caption{
Relative forward-backward production ratio $R_\mathrm{FB}$ for prompt $\Dz$ mesons over 
that for prompt $\jpsi$ mesons (left) as a function of $\pt$ 
integrated over the common rapidity range $2.5<|y^{*}|<4.0$ for $\pt<6\gevc$ and 
over $2.5<|y^{*}|<3.5$ for $6<\pt<10\gevc$;
    (right) as a function of $y^*$ integrated  up to  $\pt=10\gevc$. 
The red inner bars in the uncertainty represent the statistical 
uncertainty and the black outer bars  the quadratic sum of the 
statistical and systematic components.
} 
\label{fig:RFBD0Jpsi}
\end{figure}

In order to compare the production of 
open charm and charmonium,   the ratio of $R_\mathrm{FB}$ 
for prompt  $\jpsi$  mesons divided by $R_\mathrm{FB}$ for prompt $\Dz$ mesons is shown in Fig.~\ref{fig:RFBD0Jpsi}. 
The measurement shows that  $R_\mathrm{FB}$ has the same size 
for prompt $\Dz$ and prompt $\jpsi$ mesons 
within the uncertainties in the \lhcb kinematic range.

\section{Conclusion}

The prompt $\Dz$ production cross-section has been measured with \lhcb proton-lead collision data at
$\snn=5\Tev$.
The measurement is performed in the
range of \Dz transverse momentum $\pt<10\gevc$,
in both backward and forward collisions covering the 
ranges  $1.5 < y^{\ast} < 4.0$ and $- 5.0 < y^{\ast} < - 2.5$.
This is the first measurement of this kind down to zero transverse momentum
of the \Dz meson.
Nuclear modification factors and forward-backward production ratios are 
also measured in the same kinematic range.
Both observables are excellent probes to constrain the PDF uncertainties, 
which are currently significantly larger than 
the uncertainties on the experimental
results.
A large asymmetry in the forward-backward production is observed, which is consistent with the 
expectations from nuclear parton distribution functions, and colour glass condensate calculations 
for the forward rapidity part.
The results are found to be consistent with the theoretical predictions considered.

\section*{Acknowledgements}

\vspace{1cm}

\noindent We would like to thank Andrea Dainese, Bertrand Duclou\'e,
Jean-Philippe Lansberg and Huasheng Shao for providing the 
theoretical predictions for our measurements.
We express our gratitude to our colleagues in the CERN
accelerator departments for the excellent performance of the LHC. We
thank the technical and administrative staff at the LHCb
institutes. We acknowledge support from CERN and from the national
agencies: CAPES, CNPq, FAPERJ and FINEP (Brazil); MOST and NSFC (China);
CNRS/IN2P3 (France); BMBF, DFG and MPG (Germany); INFN (Italy); 
NWO (The Netherlands); MNiSW and NCN (Poland); MEN/IFA (Romania); 
MinES and FASO (Russia); MinECo (Spain); SNSF and SER (Switzerland); 
NASU (Ukraine); STFC (United Kingdom); NSF (USA).
We acknowledge the computing resources that are provided by CERN, IN2P3 (France), KIT and DESY (Germany), INFN (Italy), SURF (The Netherlands), PIC (Spain), GridPP (United Kingdom), RRCKI and Yandex LLC (Russia), CSCS (Switzerland), IFIN-HH (Romania), CBPF (Brazil), PL-GRID (Poland) and OSC (USA). We are indebted to the communities behind the multiple open 
source software packages on which we depend.
Individual groups or members have received support from AvH Foundation (Germany),
EPLANET, Marie Sk\l{}odowska-Curie Actions and ERC (European Union), 
Conseil G\'{e}n\'{e}ral de Haute-Savoie, Labex ENIGMASS and OCEVU, 
R\'{e}gion Auvergne (France), RFBR and Yandex LLC (Russia), GVA, XuntaGal and GENCAT (Spain), Herchel Smith Fund, The Royal Society, Royal Commission for the Exhibition of 1851 and the Leverhulme Trust (United Kingdom).

\addcontentsline{toc}{section}{References}
\setboolean{inbibliography}{true}
\bibliographystyle{LHCb}
\bibliography{main,LHCb-PAPER,LHCb-CONF,LHCb-DP,LHCb-TDR}

\newpage

 
\newpage
\centerline{\large\bf LHCb collaboration}
\begin{flushleft}
\small
R.~Aaij$^{40}$,
B.~Adeva$^{39}$,
M.~Adinolfi$^{48}$,
Z.~Ajaltouni$^{5}$,
S.~Akar$^{59}$,
J.~Albrecht$^{10}$,
F.~Alessio$^{40}$,
M.~Alexander$^{53}$,
A.~Alfonso~Albero$^{38}$,
S.~Ali$^{43}$
% ...

{\footnotesize \it
% ...
$ ^{5}$Clermont Universit{\'e}, Universit{\'e} Blaise Pascal, CNRS/IN2P3, LPC, Clermont-Ferrand, France\\
% ...
$ ^{10}$Fakult{\"a}t Physik, Technische Universit{\"a}t Dortmund, Dortmund, Germany\\
% ...
$ ^{38}$ICCUB, Universitat de Barcelona, Barcelona, Spain\\
$ ^{39}$Universidad de Santiago de Compostela, Santiago de Compostela, Spain\\
$ ^{40}$European Organization for Nuclear Research (CERN), Geneva, Switzerland\\
% ...
$ ^{43}$Nikhef National Institute for Subatomic Physics, Amsterdam, The Netherlands\\
% ...
$ ^{48}$H.H. Wills Physics Laboratory, University of Bristol, Bristol, United Kingdom\\
% ...
$ ^{53}$School of Physics and Astronomy, University of Glasgow, Glasgow, United Kingdom\\
% ...
$ ^{59}$University of Cincinnati, Cincinnati, OH, United States\\
% ...
}
\end{flushleft}

\end{document}